\documentclass[12pt,journal,cspaper,compsoc,onecolumn]{IEEEtran}
\usepackage[nocompress]{cite}
\usepackage[colorlinks,linkcolor=blue]{hyperref}
\usepackage{graphicx}
\usepackage[cmex10]{amsmath}
\usepackage{amssymb}
\usepackage{amsthm}
\usepackage{bbm}
\usepackage{array}
\usepackage{mdwmath}
\usepackage{mdwtab}
\usepackage[font=normalsize,labelfont=sf,textfont=sf]{subfig}
\usepackage{fixltx2e}
\usepackage{stfloats}
\usepackage{url}
\usepackage{color}
\usepackage{hyperref}
\usepackage{algorithmicx,algorithm}
\usepackage{algpseudocode}
\hypersetup{
    citecolor=red        % color of links to bibliography
}

\makeatletter
\newcommand{\pushright}[1]{\ifmeasuring@#1\else\omit\hfill$\displaystyle#1$\fi\ignorespaces}
\newcommand{\pushleft}[1]{\ifmeasuring@#1\else\omit$\displaystyle#1$\hfill\fi\ignorespaces}
\makeatother

\newcommand{\gcal}{\mathcal{G}}

\newcommand{\vcal}{\mathcal{V}}
\newcommand{\ecal}{\mathcal{E}}
\newcommand{\ind}{\mathbbm{1}}
\newtheorem{theorem}{Theorem}

\newtheorem{claim}{Claim}
\theoremstyle{remark}\newtheorem*{remark}{{\bf Remark}}

\begin{document}
\title{Data Requirement for Phylogenetic Inference from Multiple Loci: A New Distance Method}
%\author{Gautam~Dasarathy,~\IEEEmembership{Student Member,~IEEE,}
%        Robert~Nowak,~\IEEEmembership{Fellow,~IEEE}
%        and~Sebastien~Roch% <-this % stops a space
% without affiliation:
\author{Gautam~Dasarathy${}^\dagger$,
        Robert~Nowak${}^\dagger$,
        and~Sebastien~Roch${}^\#$\thanks{S.R. acknowledges the support of NSF grants DMS-1248176 and DMS-1149312 (CAREER), and an Alfred P. Sloan Research Fellowship.}% <-this % stops a space
        \\${}^\dagger$ Wisconsin Institutes for Discovery\\
        ${}^\#$ Department of Mathematics\\
        University of Wisconsin - Madison
%\IEEEcompsocitemizethanks{aaaa.
%aaa
%\IEEEcompsocthanksitem aaaa.}% <-this % stops a space
%}
}
%\markboth{Manuscript under preparation; Please do not distribute}%
%{Dasarathy \MakeLowercase{\textit{et al.}}: Sample Complexity of Phylogenetic Inference from Multiple Loci}
\markboth{Dasarathy \MakeLowercase{\textit{et al.}}: Data Requirement for phylogenetic inference from multiple loci}{}
\IEEEcompsoctitleabstractindextext{%
\begin{abstract}
%\boldmath
We consider the problem of estimating the evolutionary history of a set of species (phylogeny or species tree)
from several genes. It is known that the evolutionary history of individual genes (gene trees) might be topologically distinct from each other and from the underlying species tree, possibly confounding phylogenetic analysis. A further complication in practice is that one has to estimate gene trees from molecular sequences of finite length.  We provide the first full data-requirement analysis				
of a species tree reconstruction method that takes into
account estimation errors at the gene level.
Under that criterion, we also devise a novel reconstruction
algorithm that provably improves over all previous methods 
in a regime of interest. \end{abstract}
% not want either math or citations to appear in the abstract.

% Note that keywords are not normally used for peer review papers.
}

\maketitle

% is not selected <OR> if conference mode is selected - because compsoc
% conference papers position the abstract like regular (non-compsoc)
\IEEEdisplaynotcompsoctitleabstractindextext

% \fi
%\IEEEpeerreviewmaketitle

%\listoftodos

%%%%%%%%%%%%%%%%%%%%%%%%%%%%%%%%%%%%%
\section{Introduction}
\label{sec.introduction}
%%%%%%%%%%%%%%%%%%%%%%%%%%%%%%%%%%%%%

\IEEEPARstart{W}{e} consider the problem of estimating the common 
evolutionary history, more precisely the \emph{species tree}, 
of a set of $n$ species using sequence data
from multiple genes or loci. 
It is well known 
that the estimated genealogical history of 
a gene (\emph{gene tree}) may be topologically 
distinct from the species tree that encapsulates it,
possibly confounding phylogenetic analysis~\cite{maddison1997gene}.
The subject of this paper is an important source of such gene tree 
incongruence, known
as \emph{incomplete lineage sorting} (ILS), 
where two lineages fail to coalesce in their most recent
common ancestral population. That failure may lead one of the lineages to first coalesce with a more distantly related population
thereby producing a gene tree whose topology differs from the species tree that we are trying to estimate.
Several species tree reconstruction methods have recently been developed that address ILS. 
See for instance~\cite{nichols2001gene, liu2009coalescent} and references therein. 
Many such methods rely on a statistical model known as the \emph{multispecies coalescent} which, roughly speaking, generates gene trees by performing independent coalescent processes in each ancestral population and then assembling these together. This process is illustrated in
Figure~\ref{fig.mscExample2} below and explained in a little more detail in Section~\ref{sec.mscModel}.
For more background on phylogenetic inference and coalescent theory see,
e.g.,~\cite{felsenstein2004inferring,griffiths1994ancestral,semple2003phylogenetics}. 

The accuracy of multiloci reconstruction methods has been evaluated empirically,
for instance, in~\cite{leache2011accuracy,liu2009estimating}. The focus of this paper is the mathematical characterization of the performance of such
methods. Prior theoretical work has focused mainly on statistical consistency under the multispecies coalescent; see e.g., \cite{liu2009estimating,DeDeBr+:09,mossel2010incomplete,LiuYuPearl:10}. That is, assuming access to either correct gene trees or
correct pairwise distances (or coalescence times) for each gene, a method is \emph{statistically
consistent} if it is guaranteed to converge on the correct species tree as the number of genes, $m$, tends to infinity. \cite{roch2012analytical} studies the rates of convergence (in $m$) for several such methods.
%For instance, letting $f>0$ denote the smallest branch length in the species tree, in the limit $f \to 0$, it has been shown that 
%$m$ must scale as: $f^{-1}$ for the GLASS algorithm~\cite{mossel2010incomplete}, an 
%agglomerative clustering
%method using for dissimilarity
%the minimum coalescent times across genes
%for each pair of species;
%but $f^{-2}$ for the STEAC algorithm~\cite{liu2009estimating}, 
%a similar method that uses averages across genes.
For instance, letting $f>0$ denote the smallest branch length in the species tree, in the limit $f \to 0$, it was shown that the GLASS algorithm~\cite{mossel2010incomplete}, which is an agglomerative clustering method in which the dissimilarity between each pair of species is taken to be the \emph{minimum} of the coalescent times among the $m$ genes, needs the number of genes $m$ to scale as $f^{-1}$. On the other hand, $m$ needs to scale as $f^{-2}$ for the STEAC algorithm~\cite{liu2009estimating}, which is also an agglomerative clustering method which instead uses the  \emph{average} of the coalescent times across the $m$ genes as the measure of dissimilarity. 
%
%
%letting $f>0$ denote the smallest branch length in the species tree, in the limit $f\to 0$, it has been shown that $m$ must scale as $f^{-1}$ for the GLASS algorithm~\cite{mossel2010incomplete} to succeed but $f^{-2}$ for the STEAC algorithm~\cite{liu2009estimating} to succeed. 
In reality, however, one has to estimate gene trees 
and coalescent times
from finite, say, length-$k$ molecular sequences. 
Taking into account the resulting estimation errors at the gene level
is key to mathematically quantify and compare the performance of different methods (see e.g., \cite{nakhleh2013computational, yang2011fast,hahn2007bias}). Intuitively, for instance,
the ``minimum'' used in GLASS may be significantly more 
sensitive to estimation errors than the ``average'' used in STEAC. 
We make progress towards this goal by performing the first
full data requirement analysis of some species tree reconstruction
methods.

Our contribution is two-fold. 
First it is known that, in order to reconstruct a single gene tree
correctly with high probability, it is both necessary~\cite{SteelSzekely:02} and sufficient~\cite{erdos1999few} for the sequence length $k$ to scale as $f^{-2}$. Therefore, in light of this and the results in \cite{roch2012analytical}, one might expect that the total amount of data required,
$mk$, 
must scale as $f^{-3}$ and $f^{-4}$ for GLASS and STEAC
respectively.  We show that, by a crucial modification of STEAC, one obtains an algorithm that is guaranteed to reconstruct the species tree exactly with high probability as long as $m$ scales like $f^{-2}$ and $k\geq 1$. 
In particular, it suffices for the overall sample complexity, $mk$, to scale like $f^{-2}$ (which is much smaller than $f^{-3}$ and $f^{-4}$ in the regime of interest, where $f\ll1$).  
Secondly, unlike GLASS, STEAC only works under the restrictive molecular
clock assumption~\cite{semple2003phylogenetics}, where the mutation rates and population sizes are constant across the populations represented by the branches of  
the species tree.
We extend the previous data requirement result beyond the
molecular clock by devising a novel STEAC-like species tree reconstruction  algorithm which we call METAL (\underline{M}etric algorithm for \underline{E}stimation of \underline{T}rees based on \underline{A}ggregation of \underline{L}oci). This algorithm is a distance based method  where the distances are defined by concatenating the molecular sequences corresponding to all the loci (genes).

%%%%%%%%%%%%%%%%%%%%%%%%%%%%%%%%%%%%%
\section{Preliminaries and Notation}
\label{sec.preliminaries}
%%%%%%%%%%%%%%%%%%%%%%%%%%%%%%%%%%%%%
We will begin with a description of our  modeling assumptions and introduce some notation that will be used throughout the paper. 
%%%%%%%%%%%%%%%%%%%%%%%%%%%%%%%%%%%%%
\subsection{The Species Tree}
\label{sec.speciesTree}
%%%%%%%%%%%%%%%%%%%%%%%%%%%%%%%%%%%%%
\renewcommand*{\thefootnote}{\fnsymbol{footnote}}
At the heart of the model is an unknown \emph{species tree} $S = (V,E)$ which represents the evolutionary history of $n$ isolated populations; these isolated populations are represented by the size $n$ leaf set $L$ of this tree.  The goal is to learn the structure of $S$. We  assume that each branch $e\in E$ of the species tree corresponds to $t_e$ generations of evolution and we assume that each generation in this branch has a population of size $N_e$. As is standard in coalescent theory, we will assign each branch $e\in E$, a length $\tau_e>0$ in coalescent time units defined as $\tau_e\triangleq t_e/N_e$. The smallest branch length, $f\triangleq \min_e\tau_e$, will play an important role in our analysis and in particular, we will be interested in the case where $f$ is very small.  For a pair of vertices $X,Y\in V$, we will use $\pi^{S}_{XY}\subset E$ to denote the unique path connecting $X$ and $Y$ in $S$ and $\tau_{XY}$ will denote the length of this path. Notice that $\{\tau_{AB}\}_{A,B\in L}$ forms a metric on the set $L$ and such a metric that can be written as a sum of path lengths on a tree is called an \emph{additive metric}  (see e.g., \cite{semple2003phylogenetics}) with respect to that tree.
If we additionally assume that the population sizes in each branch are equal to some constant $N$, then $\{\tau_{AB}\}_{A,B\in L}$ forms an ultrametric with respect to $S$, i.e., for any three leaves $A,B,C$ such that $S$ restricted to $A,B,C$ has the topology $((A,B),C)$\footnote[1]{We will sometimes find it useful to represent trees in the so called Newick Format. For instance, the Newick representations of the trees labelled Gene 1 and Gene 2 in Figure~\ref{fig.mscExample2} are $((A,B),C)$ and $(A,(B,C))$, respectively.}, we have that 
$$\tau_{AB} \leq \tau_{AC} = \tau_{BC}.$$ We will let $\Delta\triangleq \max_{A,B\in L}\tau_{AB}$ denote the diameter of the species tree. Finally, To each branch $e\in E$, we will also associate a mutation rate, $\mu_e$ and we will let $\mu_L~\triangleq~\min_{e\in E}\mu_e$ and $\mu_U \triangleq\max_{e\in E}\mu_e$ denote the smallest and  largest mutation rates, respectively.
%%%%%%%%%%%%%%%%%%%%%%%%%%%%%%%%%%%%%
\subsection{The Multispecies Coalescent and the Gene Trees}
\label{sec.mscModel}
%%%%%%%%%%%%%%%%%%%%%%%%%%%%%%%%%%%%%
Following \cite{rannala2003bayes}, we assume that a \emph{multispecies coalescent} (MSC) process produces $m$ (independent) random genealogies $\gcal^{(1)},\gcal^{(2)},\ldots,\gcal^{(m)}$ based on $S$. These encode, say, the evolutionary history of $m$ different genes or loci on the genome and will be referred to as \emph{gene trees} henceforth.
\begin{figure}[t!]
\centering
\includegraphics[scale = 2.25]{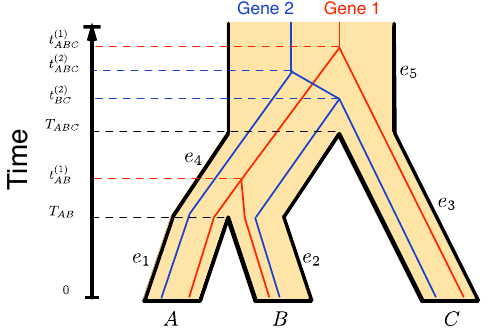}
\caption{A species tree (the thick, shaded tree) and two samples from the multispecies coalescent. Notice that while the topology of Gene~1 agrees with the species tree, the topology of Gene~2 does not.}
\label{fig.mscExample2}\vspace{-6mm}
\end{figure}

It is easier to understand the MSC constructively and in the case where the population size $N_e$ in each branch $e\in E$ is a constant $N$. Consider the 3 species example of Figure~\ref{fig.mscExample2}, where the thick, shaded tree is the species tree $S$ with edges $\{e_i\}_{i=1}^5$. As is standard in coalescent theory, we will think of time as running backwards, that is, time (in coalescent time units) starts at 0 at the leaves and increases towards the root of the tree. By $T_{AB}$ (resp. $T_{ABC}$), we mean the time when the parent population of $A$ and $B$ (resp. the parent population of $A,B,$ and $C$) branch (or speciate). Let us first consider one random draw from the MSC, i.e., the case of one particular gene,  Gene~1. $A,B,$ and $C$ each have a copy (or allele) of Gene~1 and the MSC describes the evolutionary history of the lineages corresponding to these alleles. From time $0$ until $T_{AB}$, the lineages corresponding to $A$ and $B$ are in isolated populations and hence do not ``coalesce''. However, once these lineages reach the parent population of $A$ and $B$ (represented by the branch $e_4$), they have a chance to coalesce. According to the MSC, the coalescence happens after a random time drawn according to the Exp$(1)$ distribution, that is, 
\begin{equation}
\mathbb{P}\left[t^{(1)}_{AB} - T_{AB} \geq x\right] = 1 - e^{-x},\qquad x\geq 0.
\end{equation}
Now, the coalesced $A$-$B$ lineage and the lineage corresponding to $C$ do not interact until time $T_{ABC}$, which is when they find themselves in a common population. They then coalesce at a random time $t^{(1)}_{ABC}$ which is again such that $t^{(1)}_{ABC}~-~T_{ABC}\sim$~Exp$(1)$. This gives us a random gene tree with the topology $((A,B),C)$. To contrast with this, consider the case of Gene~2. Here, the lineages corresponding to the alleles in $A$ and $B$ do not coalesce in $e_4$ (since the randomly drawn coalescence time was more than the length of $e_4$). So, at time $T_{ABC}$, there are three lineages present in the branch $e_5$. When there are multiple lineages in the same population, according to the MSC, each pair independently coalesces again after a random time period drawn according to the Exp$(1)$ distribution. In this case, the genealogies of $B$ and $C$ alleles coalesce (at time $t_{BC}^{(2)}$) before $A$ and $B$, thus giving us a second random tree with topology $(A,(B,C))$. 
%We will often make use of the fact that for any gene $\mathcal{G}^{(i)}$, the random coalescence time $t_{AB}^{(i)}$ is such that $t_{AB}^{(i)} - T_{AB}$ is distributed according to the Exp$(1)$ distribution, irrespective of which ancestral population the coalescence takes place. This follows from the following observation (called the \emph{memoryless property}): any random variable $X\sim$ Exp$(\lambda)$ satisfies
%\begin{align*}
%\mathbb{P}\left(X > t+s\mid X>t\right) = \frac{\mathbb{P}\left(X > t+s\right)}{\mathbb{P}\left(X>t\right)} =e^{-\lambda s} = \mathbb{P}(X > s)
%\end{align*}
Notice that while the genealogy (evolutionary history) of Gene~1 agrees with that of the species, the genealogy of Gene~2 does not. This is an example of incomplete lineage sorting which, as mentioned earlier,  is a fundamental road block for learning the tree of life. 

We refer the reader to \cite{rannala2003bayes} for more details on the multispecies coalescent but, we will state the model here for the sake of completeness. Before we proceed, we will record a simple fact about the exponential distribution: If $X_1,\ldots,X_p\stackrel{\rm iid}{\sim}$ Exp$(1)$, then $\min_{i\in \left\{1,\ldots,p\right\}}X_i\sim$ Exp$(p)$. This follows since  
\begin{equation}
\mathbb{P}\left(\min_{i\in \left\{1,\ldots, p\right\}}X_i\geq t\right) = \prod_{i=1}^pP(X_i\geq t) = e^{-pt}.\label{eq.minExponentials}
\end{equation}¥
The density of the likelihood of a gene tree $\mathcal{G}^{(i)} = \left(\mathcal{V}^{(i)},\mathcal{E}^{(i)}\right)$ can now be written down as follows. We will focus our attention on the branch $e\in E$ of the species tree and for the gene tree $\gcal^{(i)}$, let $I^{(i)}_e$ and $O^{(i)}_e$ be the number of lineages entering and leaving the branch $e$ respectively. For instance, consider Gene 1 in Figure~\ref{fig.mscExample2}. Here, two lineages enter the branch $e_4$  and one lineage leaves it. On the other hand, in the case of Gene~2 in Figure~\ref{fig.mscExample2}, two lineages enter the branch $e_4$ and two lineages leave it. Let $t_{e,s}^{(i)}, s = \left\{1,2,\ldots,I_e^{(i)} - O_{e}^{(i)}+1\right\}$ be the $s-$th coalescent time corresponding to $\gcal^{(i)}$ in the branch $e$. Recall that each pair of lineages in a population can coalesce at a random time drawn according to the Exp$(1)$ distribution  independently of each other. Therefore, after the $(s-1)$-th coalescent event at time $t_{e,s-1}^{(i)}$, there are $I_e^{(i)} - s+1$ surviving lineages in branch $e$ and the likelihood that the $s-$th coalescence time in branch $e$ is $t_{e,s}^{(i)}$ corresponds to the event that the minimum of ${I_e^{(i)} - s+1\choose 2}$ random variables distributed according to Exp$(1)$ has the value $t_{e,s}^{(i)} - t_{e,s-1}^{(i)}$. Therefore using \eqref{eq.minExponentials}, the density of the likelihood of $\gcal^{(i)}$ can be written as 
\begin{equation}
\prod_{e\in E}\prod_{s=1}^{I^{(i)}_e-O_e^{(i)}+1}\exp\left\{-{I_e^{(i)} -s+1\choose 2}\left[t_{e,s}^{(i)} - t_{e,s-1}^{(i)}\right]\right\},
\end{equation}¥
where, for convenience, we let $t^{(i)}_{e,0}$ and $t^{(i)}_{e,I^{(i)}_{e}-O^{(i)}_{e}+1}$ be respectively the divergence times of the population in $e$ and of its parent population. 
%\begin{align*}
%\prod_{{e\in E}} 
%\exp\left(-\binom{O^{(i)}_{e}}{2} 
%\left[t^{(i)}_{e, O^{(i)}_{e}+1} - t^{(i)}_{e, O^{(i)}_{e}}\right]\right)\times\prod_{s=1}^{I^{(i)}_{e}-O^{(i)}_{e}}
%\exp\left(-\binom{s}{2} 
%\left[t^{(i)}_{e, s} - t^{(i)}_{e, s-1}\right]\right),
%\end{align*}
%where, for locus (or gene) $i$ and branch $e$, $I^{(i)}_{e}$ is the number of lineages entering $e$, $O^{(i)}_{e}$ is the number of lineages exiting $e$, and
%$t^{(i)}_{e,s}$ is the $s$-th coalescence time in $e$;

\renewcommand*{\thefootnote}{\arabic{footnote}}

%%%%%%%%%%%%%%%%%%%%%%%%%%%%%%%%%%%%%
\subsection{Observation Model and The Inference Problem}
\label{sec.observationModel}
Much of the prior work on understanding the theoretical complexity of learning species trees from multiple loci (or gene trees) has focused on the case where exact gene trees are available. However, in reality one needs to estimate these gene trees from molecular sequences and indeed there has been a recent thrust towards understanding the effect of errors in estimating the gene trees (see e.g., \cite{nakhleh2013computational,yang2011fast,hahn2007bias}). Our approach will be to take this error into account explicitly and in fact bypass the reconstruction of gene trees altogether. 

We model the sample generation process according to the standard Jukes-Cantor (JC) model (see e.g., \cite{semple2003phylogenetics}). That is, given a gene tree $\gcal = (\vcal,\ecal)$, we will associate to each $\tilde{e}\in \ecal$, a probability $p_{\tilde{e}}$ (whose dependence on the length of $\tilde{e}$ we will make explicit below). Then, the JC model assigns a character from $\{{\tt A,T,G,C}\}$ uniformly at random to the root of $\mathcal{G}$. Moving away from the root, with probability $p_{\tilde{e}}$, each edge $\tilde{e}$ changes the state of its ancestor to one of the other three, chosen uniformly at random. The states at the leaves of $\mathcal{G}$ are assembled into a length $n$ vector to get the first sample; this process is repeated $k$ times to generate the data set. Notice that $k$ models the number of sites or the sequence length of each gene.  

Now, we will define $p_{\tilde{e}}$. To each edge $\tilde{e}$ of the random gene tree $\gcal$ is associated a random length $\sigma_{\tilde{e}}$ according to the MSC. Also, given an edge $e\in E$ of the species tree, we will write $\sigma_{e\cap \tilde{e}}$ to denote the length of the portion $\tilde{e}$ that overlaps with $e$. This lets us define the effective (mutation rate adjusted) branch lengths, $\delta_{\tilde{e}} = \sum_{e\in E}\mu_e\sigma_{e\cap \tilde{e}}$. As before, for any two vertices $X,Y\in\vcal$, $\pi_{XY}^\gcal$ denotes the path joining $X$ and $Y$ in $\gcal$ and $\sigma_{XY}$ (resp. $\delta_{XY}$) denotes the length of this path under $\sigma$ (resp. under $\delta$). Now, for an edge $\tilde{e}\in \ecal$, we define $p_{\tilde{e}} \triangleq \frac{3}{4}(1 - e^{-\frac{4}{3}\delta_{\tilde{e}}})$. Notice that this definition implies that the probability $p_{XY}$ of disagreement between the characters at vertices $X$ and $Y$ satisfies, $p_{XY} =\frac{3}{4}(1 - e^{-\frac{4}{3}\delta_{XY}})$.

%\textcolor{red}{The Jukes-Cantor model was chosen because it lends itself to easy presentation. since the methods we develop are ``distance-based'', all our results generalize to the more realistic GTR model (see e.g., \cite{semple2003phylogenetics}) say, using spectral methods.}

The goal then, is  to learn the structure of $S$ given the data $\left\{\chi^{ij}\right\}_{i\in [m],j\in [k]}$ which is an $n\times m\times k$ array composed of the characters $\{{\tt A,T,G,C}\}$, where $\{\chi^{ij}\}_{j\in [k]}$ is the data generated from the random gene tree $\gcal^{(i)}$ according to the Jukes-Cantor model.  

The Jukes-Cantor model was chosen because it lends itself to easy presentation. Since the techniques developed here are \emph{distance-based}, all our results can be generalized to the more realistic Generalized Time-Reversible (GTR) model \cite{tavare1986some} using spectral techniques as in \cite{roch2010toward, mossel2003information}.
%%%%%%%%%%%%%%%%%%%%%%%%%%%%%%%%%%%%%
\section{Main Results}
\label{sec.mainResults}
We now state the main results of the paper. First, we will deal with the case where the strong molecular clock \cite{semple2003phylogenetics} assumption holds. We will then turn our attention to the more general case that does away with this assumption. 
\subsection{The Molecular Clock Assumption Holds}
\label{sec.ultrametric}
Assuming that the molecular clock hypothesis holds is often unrealistic; it is equivalent to believing that all extant and ancestral populations have the same population size and that the mutations happen at the same rate through time and across populations. It has however proven to be a useful abstraction for developing powerful methods. In our setting, this is equivalent to assuming that for all $e\in E$, $\mu_e = \mu >0$,  and $N_e = N$, both constants independent of $e$. 

In order to infer the species tree from samples, we will begin by defining a distance measure on the leaves. For each pair of leaves $A,B\in L$, we define 
\begin{equation}
\widehat{p}_{AB} = \frac{1}{mk}\sum_{i\in[m],j\in[k]}\ind\{\chi^{ij}_{A}\neq\chi^{ij}_B\}\label{eq.SampleUltrametricDistances},
\end{equation}
which can be thought of as the normalized hamming distance between the concatenated molecular sequences corresponding to species $A$ and $B$. Our first result, which is proved in Appendix~\ref{sec.proofOfTheorem1}, is that, in expectation, $\left\{\widehat{p}_{AB}\right\}_{A,B\in L}$ is not only a metric on $L$, but is in fact an ultrametric with respect to $S$. 
\renewcommand*{\thefootnote}{\fnsymbol{footnote}}
\begin{theorem}
\label{thm.ExpectationUltrametric}
$\{\mathbb{E}\left[\widehat{p}_{AB}\right]\}_{A,B\in L}$ forms an ultrametric with respect to the true species tree $S$. In fact, for any triple $A,B,C\in L$ with the topology $((A,B),C)$ in $S$, we have
\begin{equation}
\mathbb{E}\left[\widehat{p}_{AC}\right] = \mathbb{E}\left[\widehat{p}_{BC}\right] > \mathbb{E}\left[\widehat{p}_{AB}\right] +  \frac{3e^{-\frac{4}{3}\mu \tau_{AC}}\mu }{8\mu + 3} \,f\label{eq.ExpectationUltrametricDistances}.
\end{equation}
\end{theorem}
\renewcommand*{\thefootnote}{\arabic{footnote}} 
%follows from the observation that, by definition, we have $\mathbb{E}\left[\widehat{p}_{AC}\right]= \mathbb{E}\left[\frac{3}{4}\left(1 - e^{-\frac{4}{3}{\delta_{AC}}}\right)\right]$, where $\delta_{AC}$ is the random gene tree distance that satisfies $\delta_{AC} = \mu\tau_{AC} + 2\mu Z$ with $Z\sim \mbox{Exp}(1)$. We refer the reader to \cite{isit14_appendix} for the exact details. 
This result inspires the following procedure for reconstructing $S$: Use $\{\widehat{p}_{AB}\}_{A,B\in L}$ as a dissimilarity measure for $L$ and use a standard algorithm that accepts a dissimilarity measure and returns an ultrametric tree (see e.g., \cite{felsenstein2004inferring,semple2003phylogenetics} for background on distance based methods). For the sake of simplicity, we may assume that we use the UPGMA algorithm\cite{sokal1958statistical}, the standard method for bottom-up agglomerative clustering, in order to produce an ultrametric tree. Then, recalling that $\mu$ denotes the (common) mutation rate across the populations represented by the species tree $S$, and $\Delta$ denotes diameter of $S$, we have the following performance guarantee.
\begin{theorem}
\label{thm.ultrametricSampleComplexity}
Given an $\epsilon > 0$, using UPGMA on $L$ with the dissimilarity measure $\{\widehat{p}_{AB}\}_{A,B\in L}$ results in the correct tree $S$ being output with probability no less than $1-\epsilon$ as long as the number of genes  $m$, and the sequence length $k$ satisfy
\begin{equation}
m \geq C_1(\mu,\Delta,n,\epsilon) \times  f^{-2}\;\mbox{ and }\; k \geq 1\label{eq.ultrametricSampleComplexity},
\end{equation}
where $C_1(\mu,\Delta,n,\epsilon) = \frac{16\, e^{\frac{8}{3}\mu \Delta}(8\mu+3)^2}{9\mu^2}\log\left(\frac{8{n\choose 3}}{\epsilon}\right)$. 
\end{theorem}
%\begin{remark}We know from \cite{erdos1999few} that the diameter $\Delta$ in the above expression can be replaced by the (often much smaller) ``depth'' of the tree\footnote{Recall that the depth of an edge $e$ is the length (under $\tau$) of the shortest path between two leaves crossing $e$. The depth of a tree is the maximum edge depth.} by using an algorithm that only relies on distances that are ``close enough''.
%\end{remark}

Theorem~\ref{thm.ultrametricSampleComplexity}, which is proved in Appendix~\ref{sec.proofOfTheorem2}, tells us that the above procedure succeeds with high probability as long as we get molecular sequences of length at least one from at least $\mathcal{O}(f^{-2})$ genes. That is, a total sequence length of $mk=\mathcal{O}(f^{-2})$ suffices for reliable learning. 

Notice that the procedure we propose is similar to the STEAC algorithm \cite{liu2009estimating} except instead of using the average coalescent time as the distance measure, we use \eqref{eq.SampleUltrametricDistances}, which can be considered as the normalized hamming distance. It turns out that this modification is crucial to obtaining our improved sample complexity result. 

\subsection{The Molecular Clock Assumption Does Not Hold}
\label{sec.noMolecularClock}
We will now consider the more general case where the strong molecular clock assumption does not hold. That is, we will assume that each branch $e$ of the species tree has a (possibly) distinct mutation rate $\mu_e$ and population size $N_e$.   

First, we observe that $\{\mathbb{E}[\widehat{p}_{AB}]\}_{A,B\in L}$ as defined above is no longer an ultrametric with respect to $S$ and therefore, the above procedure (and for a similar reason, the STEAC algorithm) cannot be used to recover the species tree. In such situations, one usually turns to distance methods that rely on the 4-point condition (see e.g., \cite{semple2003phylogenetics}). However, it is not immediately clear how to define a metric that satisfies the 4-point condition in our setting. Our next result, which is arguably the most important contribution of this paper, shows that this can be done. As before, we will first consider an idealized measure of dissimilarity as follows: 
%Towards this end, for $A,B\in L$, we define the following measure of dissimilarity 
$$d_{AB} = -\frac{3}{4}\log\left(1 - \frac{4}{3}\mathbb{E}\left[\widehat{p}_{AB}\right]\right), A,B\in L,$$
where $\widehat{p}_{AB}$ is as defined in \eqref{eq.SampleUltrametricDistances}. 
Our next result, which parallels Theorem~\ref{thm.ExpectationUltrametric}, shows that this ``idealized'' dissimilarity measure is actually an  \emph{additive metric}  with respect to $S$. Recall that this means that the four point condition holds, i.e., for a quadruple of leaves $A,B,C,D$ that are such that the topology of $S$ restricted to these 4 leaves is $((A,B),(C,D))$ or $(((A,B),C),D)$, the above distances satisfy $$d_{AB} + d_{CD} \leq d_{AC} + d_{BD} = d_{AD} + d_{BC}.$$ 
See \cite{semple2003phylogenetics}, for instance, for more information about tree metrics. 
%
%
%Towards this end, we propose the following \emph{corrected} measure of dissimilarity (with $\widehat{p}_{AB}$ as defined in \eqref{eq.SampleUltrametricDistances})
%\begin{align}
%\widehat{d}_{AB} &\triangleq-\frac{3}{4}\log\left(1 - \frac{4}{3}\,\widehat{p}_{AB}\right)\label{eq.nonultrametricSampleDistances}%= -\frac{1}{2}\log\left(1 - 2\widehat{p}_{AB}\right)
%\end{align}
%\renewcommand*{\thefootnote}{\fnsymbol{footnote}}
%As before, we will first show that if, for $A,B\in L$, we define $d_{AB} = -\frac{3}{4}\log\left(1 - \frac{4}{3}\mathbb{E}\left[\widehat{p}_{AB}\right]\right)$, these distances actually form an (see e.g., \cite{semple2003phylogenetics}) with respect to $S$. 
\begin{theorem}
\label{thm.nonultrametricDistances}
The set of dissimilarities $\{d_{AB}\}_{A,B\in L}$ forms an additive metric with respect to $S$. In fact, suppose the leaves $A,B,C,D \in L$ are such that either $((A,B),(C,D))$ or $(((A,B),C),D)$ holds with respect to $S$, then 
\begin{align}
d_{AC} + d_{BD} &= d_{AD} + d_{BC}> d_{AB} + d_{CD} + \alpha_{\rm add},
\end{align}
where $\alpha_{\rm add} = \frac{3}{4}\log\left(\frac{8}{3}\mu_L(1-e^{-f}) + 1\right)>0$ and $\mu_L\triangleq \min_{e\in E}\mu_e$ is the smallest mutations rate, as defined in Section~\ref{sec.speciesTree}.   
\end{theorem}

It is somewhat surprising that this result is true. It tells us that if one ignores the fact that there are multiple loci and pretends as though all samples came from a single gene tree, then the gene tree estimated from this ``concatenated molecular sequence'' has the same topology as $S$. Furthermore, this result is also  interesting since phylogenetic mixtures are known to cause problems for distance-based methods \cite{Steel2009467}. We prove Theorem~\ref{thm.nonultrametricDistances} in Appendix~\ref{sec.proofOfTheorem3}. 

In light of this, we propose the following algorithm to reconstruct $S$. First, we define the following sample-based  \emph{corrected} measure of dissimilarity (with $\widehat{p}_{AB}$ as defined in \eqref{eq.SampleUltrametricDistances})
\begin{align}
\widehat{d}_{AB} &\triangleq-\frac{3}{4}\log\left(1 - \frac{4}{3}\,\widehat{p}_{AB}\right).\label{eq.nonultrametricSampleDistances}%= -\frac{1}{2}\log\left(1 - 2\widehat{p}_{AB}\right)
\end{align}
Now, use any quartet-test based algorithm (like Neighbor Joining \cite{saitou1987neighbor}) which returns an additive tree using $\{\widehat{d}_{AB}\}_{A,B\in L}$ defined as in \eqref{eq.nonultrametricSampleDistances} as the input dissimilarity measure. We call this algorithm METAL (for \underline{M}etric algorithm for \underline{E}stimation of \underline{T}rees based on \underline{A}ggregation of \underline{L}oci).

%We call this algorithm SLIC, which stands for \emph{\underline{S}pecies tree \underline{L}earn\underline{I}ng based on \underline{C}oncatenation}. 
Recall that $\mu_U$ and  $\mu_L$ are respectively the maximum and minimum mutation rates, and $\Delta$ is the diameter of the species tree $S$ (c.f. Section~\ref{sec.speciesTree}). We then have the following result. 

\begin{theorem}
\label{thm.nonultrametricSampleComplexity}
For any $\epsilon >0$, METAL succeeds in reconstructing (the unrooted version of) $S$ with probability at least $1 - \epsilon$ as long as $m$ and $k$ satisfy
\begin{align}
k\geq 1\,\mbox{and}\,\,m &\geq \frac{e^{\frac{8\mu_U \Delta}{3}}({8}\mu_U+3)^2(24+8\alpha_{\rm add})^2 }{162\alpha_{\rm add}^2}\log\left(\frac{16{n\choose 4}}{\epsilon}\right)\label{eq.thm4SampleComplexity}
\end{align}
where $\alpha_{\rm add} = \frac{3}{4}\log\left(\frac{8}{3}\mu_L(1-e^{-f}) + 1\right)$. 

In the limit as $f\to 0$, the right side above approaches $$C_2(\mu_U,\mu_L,\Delta,n,\epsilon)~\times~f^{-2}, \mbox{ where } C_2(\mu_U,\mu_L,\Delta,n,\epsilon) = \frac{8e^{\frac{8\mu_U \Delta}{3}}(8\mu_U+3)^2 }{9\mu_L^2}\log\left(\frac{16{n\choose 3}}{\epsilon}\right).$$ 
\end{theorem}

\begin{remark} Following \cite{erdos1999few}, the diameter $\Delta$ can be replaced by the (often much smaller) \emph{depth}\footnote{The depth of an edge $e$ is the length (under $\tau$) of the shortest path between two leaves crossing $e$; the depth of a tree is the maximum edge depth.}  of the tree by employing a distance method that uses only those distances that are ``small enough''.
\end{remark}
\renewcommand*{\thefootnote}{\arabic{footnote}}

We prove Theorem~\ref{thm.nonultrametricSampleComplexity} using arguments that are similar in spirit to those in the proof of Theorem~\ref{thm.ultrametricSampleComplexity}. We refer the reader to Appendix~\ref{sec.proofOfTheorem4} for the exact details. 

Theorem~\ref{thm.nonultrametricSampleComplexity} tells us that as long as $m$ scales like $\mathcal{O}(f^{-2})$ and $k\geq 1$, the species tree can be reconstructed (upto the location of the root) reliably. 
%In \cite{mossel2010incomplete}, the authors suggest how the GLASS algorithm can be adapted to the non-ultrametric case. Therefore, one can again compare the results of this paper to the sample complexity of GLASS which would require that $m$ and $k$ scale as $\Omega(f^{-1})$ and $\Omega(f^{-2})$, respect 
It should be noted here that we assume that for each population/branch $e\in E$, the mutation rate $\mu_e$ is constant across gene trees; generalizing this analysis to the case where the mutation rates are allowed to change is an interesting avenue for future work.
\section{Discussion}
Irrespective of the sequence length $k$ of each gene, the number of genes $m$ required needs to satisfy $m\in \Omega(f^{-1})$ for consistent species tree estimation. To see this, consider the species tree in Figure~\ref{fig.mscExample2}. Given $m$ gene trees drawn  according to the MSC based on this species tree, the probability that none of them  have a coalescent event in branch $e_4$ is given by $e^{-m\tau_{e_4}}$ (this is the probability that $m$ independent exponentials are bigger than $\tau_{e_4}$). Therefore, if $m < \tau_{e_4}^{-1}$, then with probability greater than $e^{-1}$, none of the $m$ the gene trees have a coalescence event in $e_4$, that is, there is no evidence for the existence of this branch from the sample. This argument can also be formalized by observing that any algorithm that is able to estimate $S$ reliably should be able to perform a reliable hypothesis test between two shifted exponential distributions. Therefore, this result follows from the fact that $D_{\rm KL}\left(p(x;\tau_{AB}+f)\middle\|p(x;\tau_{AB})\right) = f$, where $p(x;a) = e^{-(x-a)}\ind\left\{x\geq a\right\}$ and $D_{KL}\left(\cdot\|\cdot\right)$ is the Kullback-Liebler divergence \cite{cover2012elements}. 

On the other hand, we know from \cite{SteelSzekely:02} that even without the confounding effect of the multispecies coalescent, a total sequence length ($m\times k$) of at least $\Omega(f^{-2})$ is needed for consistent estimation. These two together imply that there is a constant $C>0$ such that $m$ needs to satisfy the following for consistent estimation of the species tree 
\begin{equation}
m \geq C\max\left\{f^{-1},\frac{f^{-2}}{k}\right\} \label{eq.lowerBound}.
\end{equation}

As mentioned earlier,  the results in this paper show that $m\in \mathcal{O}(f^{-2})$ is achievable irrespective of the value of $k$, i.e., in particular, a total data set size of $mk\in \mathcal{O}(f^{-2})$ is achievable. Prior to this, to the best of our knowledge, the best complexity bounds were provably attained by GLASS \cite{mossel2010incomplete} (as shown in \cite{roch2012analytical}) which requires that $m\geq \mathcal{O}(f^{-1})$ and $k\geq \mathcal{O}(f^{-2})$, i.e., a total data set size of $mk\in \mathcal{O}(f^{-3})$. 

This raises two very interesting open questions. (A) What is the precise tradeoff between $m$ and $k$ for reliable recovery of $S$ and in particular, is it possible to devise an algorithm that recovers $S$ given $m\in o(f^{-2})$ when the sequence length, $k$, is moderate, say, $\mathcal{O}(f^{-1})$? (B) Is there a procedure that attains all points (values of $m$ and $k$) in this tradeoff, as opposed to the current situation where it appears as though GLASS meets the lower bounds for large $k$ and METAL meets the  lower bound for small $k$?
%\pagebreak
\bibliographystyle{ieeetr}
\bibliography{refs}
\appendices
\newtheorem*{theorem1}{Theorem 1}
\newtheorem*{theorem2}{Theorem 2}
\newtheorem*{theorem3}{Theorem 3}
\newtheorem*{theorem4}{Theorem 4}
\newtheorem*{theorem5}{Theorem 5}
%%%%%%%%%%%%%%%%%%%%%%%%%%%%%%%%%%%%%
\section{Proof of Theorem ~\ref{thm.ExpectationUltrametric}}
\label{sec.proofOfTheorem1}
%%%%%%%%%%%%%%%%%%%%%%%%%%%%%%%%%%%%%
Recall that for any pair of leaves $A,B\in L$, we define 
\begin{equation}
\widehat{p}_{AB} = \frac{1}{mk}\sum_{i\in[m],j\in[k]}\ind\{\chi^{ij}_{A}\neq\chi^{ij}_B\}.
\end{equation}
\renewcommand*{\thefootnote}{\fnsymbol{footnote}}
\begin{theorem1}
%\label{thm.ExpectationUltrametric}
$\{\mathbb{E}\left[\widehat{p}_{AB}\right]\}_{A,B\in L}$\footnote[2]{Unless otherwise noted, expectations will be with respect all the randomness present.} forms an ultrametric with respect to the true species tree $S$. In fact, for any triple $A,B,C\in L$ with the topology $((A,B),C)$ in $S$, we have
\begin{equation}
\mathbb{E}\left[\widehat{p}_{AC}\right] = \mathbb{E}\left[\widehat{p}_{BC}\right] > \mathbb{E}\left[\widehat{p}_{AB}\right] +  \frac{3e^{-\frac{4}{3}\mu \tau_{AC}}\mu f}{8\mu + 3}\label{eq.ExpectationUltrametricDistances}.
\end{equation}¥
\end{theorem1}
\renewcommand*{\thefootnote}{\arabic{footnote}}
\begin{IEEEproof}
Suppose that $A,B,C\in L$ are three arbitrary leaves of the species tree with the topology $((A,B),C)$. By definition, we have that 
\begin{align*}
\mathbb{E}\left[\widehat{p}_{AC}\right] &= \mathbb{E}\left[\frac{3}{4}\left(1-e^{-\frac{4}{3}\delta_{AC}}\right)\right],
\end{align*}
where $\delta_{AC}$ is the distance between $A$ and $C$ on a random gene tree drawn according to the MSC. Notice that it satisfies $\delta_{AC} = \mu\tau_{AC} + 2\mu Z$ with $Z\sim \mbox{Exp}(1)$. Therefore, we have 

\begin{align*}
\mathbb{E}\left[\widehat{p}_{AC}\right] - \mathbb{E}\left[\widehat{p}_{AB}\right]&= -\frac{3}{4}e^{-\frac{4}{3}\mu\tau_{AC}}\mathbb{E}\left[e^{-\frac{8}{3}\mu Z}\right] + -\frac{3}{4}e^{-\frac{4}{3}\mu\tau_{AB}}\mathbb{E}\left[e^{-\frac{8}{3}\mu Z}\right]\\
&\stackrel{(a)}{=}\frac{3\left(e^{-\frac{4}{3}\mu\tau_{AB}} - e^{-\frac{4}{3}\mu\tau_{AC}}\right)}{4(\frac{8}{3}\mu+1)}\\
&\stackrel{(b)}{\geq} \frac{3e^{-\frac{4}{3}\mu\tau_{AC}}\mu f}{(8\mu+3)},
\end{align*}
where $(a)$ follows from the fact that if $X\sim $ Exp$(1)$, for any $\alpha>0$, $\mathbb{E}[e^{-\alpha X}] = (\alpha+1)^{-1}$ and $(b)$ follows from observing that for any $\alpha>0$ and $x<y$, we have 
\begin{align*}
\frac{e^{-\alpha x}}{\alpha} - \frac{e^{-\alpha y}}{\alpha} = \int_x^ye^{-\alpha t}\,dt \geq (y-x)e^{-\alpha y}
\end{align*}
Proceeding similarly, It can be seen that $\mathbb{E}\left[\widehat{p}_{AC}\right] = \mathbb{E}\left[\widehat{p}_{BC}\right]$. This concludes the proof. 
\end{IEEEproof}

%%%%%%%%%%%%%%%%%%%%%%%%%%%%%%%%%%%%%
\section{Proof of Theorem~\ref{thm.ultrametricSampleComplexity}}
\label{sec.proofOfTheorem2}
%%%%%%%%%%%%%%%%%%%%%%%%%%%%%%%%%%%%%
We now prove Theorem~\ref{thm.ultrametricSampleComplexity} which guarantees that $S$ can be reliably recovered by using a standard distance-based algorithm like UPGMA or bottom-up agglomerative clustering with $\{\widehat{p}_{AB}\}_{A,B\in L}$ as a dissimilarity measure for $L$. 
%\begin{theorem2}
%Given an $\epsilon > 0$, using UPGMA on $L$ with the dissimilarity measure $\{\widehat{p}_{AB}\}_{A,B\in L}$ results in the correct tree $S$ being output with probability no less than $1-\epsilon$ as long as the number of gene trees $m$, and the number of samples per gene tree $k$ satisfy
%\begin{equation}
%m \geq \frac{16\, e^{\frac{8}{3}\mu \Delta}(8\mu+3)^2}{9\mu^2f^2}\log\left(\frac{4{n\choose 3}}{\epsilon}\right)\;\mbox{ and }\; k \geq 1\label{eq.ultrametricSampleComplexity}
%\end{equation}
%\end{theorem2}
\begin{theorem2}
%\label{thm.ultrametricSampleComplexity}
Given an $\epsilon > 0$, using UPGMA on $L$ with the dissimilarity measure $\{\widehat{p}_{AB}\}_{A,B\in L}$ results in the correct tree $S$ being output with probability no less than $1-\epsilon$ as long as the number of genes  $m$, and the sequence length $k$ satisfy
\begin{equation}
m \geq C_1(\mu,\Delta,n,\epsilon) \times  f^{-2}\;\mbox{ and }\; k \geq 1, \label{eq.ultrametricSampleComplexity2}
\end{equation}
where $C_1(\mu,\Delta,n,\epsilon) = \frac{16\, e^{\frac{8}{3}\mu \Delta}(8\mu+3)^2}{9\mu^2}\log\left(\frac{8{n\choose 3}}{\epsilon}\right)$. 
\end{theorem2}

\begin{IEEEproof}
Recall that the algorithm we propose to recover the tree uses $\{\hat{p}_{AB}\}_{A,B\in L}$ as a dissimilarity measure and uses an agglomerative clustering algorithm. Therefore, this procedure errs if for any triple of leaves $A,B,C$ which have the topology $((A,B),C)$ with respect to $S$, either $\widehat{p}_{AB} > \widehat{p}_{AC}$ or $\widehat{p}_{AB} > \widehat{p}_{BC}$. Letting ${L\choose 3}$ denote the set of all unordered triples in $L$, we can use the union bound and over-estimate the error as follows  
\begin{align}
\mathbb{P}\left[\mbox{Error}\right] %&= \mathbb{P}\left[\exists \mbox{a triple $A,B,C$ such that the topology is wrongly learnt}\right]\nonumber\\
& = \mathbb{P}\left[\bigcup_{((A,B),C)\in {L\choose 3}}\Big\{\mbox{The triple $((A,B),C)$ is such that $\widehat{p}_{AB} > \widehat{p}_{AC}$ or $\widehat{p}_{AB} > \widehat{p}_{BC}$}\Big\}\right]\nonumber\\
&\leq \sum_{((A,B),C)\in {L\choose 3}}\mathbb{P}\left[\widehat{p}_{AB} > \widehat{p}_{AC}\right] + \mathbb{P}\left[\widehat{p}_{AB} > \widehat{p}_{BC}\right].\label{eq.probError}
\end{align}
We will now upper bound the term $\mathbb{P}\left[\widehat{p}_{AB} > \widehat{p}_{AC}\right]$, the other term will satisfy the same upper bound. Defining $\alpha_{\rm um} = \frac{3e^{-\frac{4}{3}\Delta}\mu f}{(8\mu+3)}$, for an arbitrary triple $((A,B),C)$ we have 
\begin{align}
\mathbb{P}\left[\widehat{p}_{AB} - \widehat{p}_{AC}>0\right] &= \mathbb{P}\left[\widehat{p}_{AB} -\mathbb{E}\left[\widehat{p}_{AB}\right] - \widehat{p}_{AC} + \mathbb{E}\left[\widehat{p}_{AC}\right]>\mathbb{E}\left[\widehat{p}_{AC}\right] - \mathbb{E}\left[\widehat{p}_{AB}\right]\right]\nonumber\\
&\stackrel{(a)}{\leq} \mathbb{P}\left[\widehat{p}_{AB} -\mathbb{E}\left[\widehat{p}_{AB}\right] - \widehat{p}_{AC} + \mathbb{E}\left[\widehat{p}_{AC}\right]>\alpha_{\rm um}\right]\nonumber\\ &\leq \mathbb{P}\left[\widehat{p}_{AB} -\mathbb{E}\left[\widehat{p}_{AB}\right] > \frac{\alpha_{\rm um}}{2}\right] + \mathbb{P}\left[\mathbb{E}\left[\widehat{p}_{AC}\right] - \widehat{p}_{AC}> \frac{\alpha_{\rm um}}{2}\right],\label{eq.probErrorABC}
\end{align}
where $(a)$ follows from Theorem~\ref{thm.ExpectationUltrametric}.
Let us first look at the first term in \eqref{eq.probErrorABC}. The second one will follow similarly. 
\begin{align}
\mathbb{P}&\left[\widehat{p}_{AB} - \mathbb{E}[p_{AB}] > \alpha_{\rm um}/2\right]\nonumber \\&\stackrel{(a)}{=} \mathbb{E}\left[\mathbb{P}\left(\widehat{p}_{AB} - \frac{1}{m}\sum_{i\in [m]}p^{(i)}_{AB} + \frac{1}{m}\sum_{i\in [m]}p^{(i)}_{AB}- \mathbb{E}\left[\widehat{p}_{AB}\right] > \frac{\alpha_{\rm um}}{2}\middle|\{\delta_{AB}^{(i)}\}_{i\in [m]}\right)\right]\nonumber\\
&\leq \mathbb{E}\left[\mathbb{P}\left(\widehat{p}_{AB} - \frac{1}{m}\sum_{i\in [m]}p^{(i)}_{AB} > \frac{\alpha_{\rm um}}{4}\middle|\{\delta_{AB}^{(i)}\}_{i\in [m]}\right) + \mathbb{P}\left(\frac{1}{m}\sum_{i\in [m]}p^{(i)}_{AB} - \mathbb{E}\left[\widehat{p}_{AB}\right] > \frac{\alpha_{\rm um}}{4}\middle|\{\delta_{AB}^{(i)}\}_{i\in [m]}\right)\right]\nonumber\\
&=\mathbb{E}\left[\mathbb{P}\left(\widehat{p}_{AB} - \frac{1}{m}\sum_{i\in [m]}p^{(i)}_{AB} > \frac{\alpha_{\rm um}}{4}\middle|\{\delta_{AB}^{(i)}\}_{i\in [m]}\right) \right] + \mathbb{P}\left(\frac{1}{m}\sum_{i\in [m]}p^{(i)}_{AB} - \mathbb{E}\left[\widehat{p}_{AB}\right] > \frac{\alpha_{\rm um}}{4}\right).
%&\leq e^{-mk\alpha_{\rm um}^2/16} + e^{-m\alpha_{\rm um}^2/16}
\end{align}
In $(a)$, we condition on $\{\delta_{AB}^{(i)}\}_{i\in [m]}$, where $\delta_{AB}^{(i)}$, as before, is the random distance between the leaves $A$ and $B$ on the gene tree $\mathcal{G}^{(i)}$. We then add and subtract $\frac{1}{m}\sum_{i\in [m]}p^{(i)}_{AB}$, where $p_{AB}^{(i)}\triangleq \frac{3}{4}\left(1 - e^{-\frac{4}{3}\delta^{(i)}_{AB}} \right)$. The next inequality follows from a union bound. The two terms in the last equation can now be upper bounded using Hoeffding's inequality: 
\begin{align}
\mathbb{E}\left[\mathbb{P}\left[\frac{1}{mk}\sum_{i=1}^m\sum_{j=1}^kX^{ij}_{AB} - \frac{1}{m}\sum_{i=1}^mp^{(i)}_{AB} > \frac{\alpha_{\rm um}}{4}\middle|\left\{d_{AB}^{(i)}\right\}\right]\right] &\leq e^{-mk\alpha_{\rm um}^2/16}\label{eq.thm2Hoeffding1}.\\
\mathbb{P}\left(\frac{1}{m}\sum_{i\in [m]}p^{(i)}_{AB} - \mathbb{E}\left[\widehat{p}_{AB}\right] > \frac{\alpha_{\rm um}}{4}\right) &\leq e^{-m\alpha_{\rm um}^2/16}\label{eq.thm2Hoeffding2}.
\end{align} 
These inequalities follow since $\mathbb{E}\left[X^{ij}_{AB}\middle | \delta^{(i)}_{AB}\right] = p^{(i)}_{AB}$ and $\mathbb{E}\left[p^{(i)}_{AB}\right] = \mathbb{E}\left[\widehat{p}_{AB}\right]$. 

Substituting these in \eqref{eq.probError}, we have 
\begin{align*}
\mathbb{P}\left[\mbox{Error}\right] &\leq \sum_{((AB)C)\in{L\choose 3}}\mathbb{P}\left[\widehat{p}_{AB} > \widehat{p}_{AC}\right] + \mathbb{P}\left[\widehat{p}_{AB} > \widehat{p}_{BC}\right]\\
&\leq \sum_{((AB)C)\in{L\choose 3}}4\left(e^{-mk\alpha_{\rm um}^2/16} + e^{-m\alpha_{\rm um}^2/16}\right)\\
&\leq {n\choose 3} 4\left(e^{-mk\alpha_{\rm um}^2/16} + e^{-m\alpha_{\rm um}^2/16}\right)
\end{align*}
Therefore, the  probability of error can be made less than $\epsilon$ if we pick $m$ and $k$ as shown in \eqref{eq.ultrametricSampleComplexity} or \eqref{eq.ultrametricSampleComplexity2}.  
\end{IEEEproof}
%%%%%%%%%%%%%%%%%%%%%%%%%%%%%%%%%%%%%
\section{Proof of Theorem~\ref{thm.nonultrametricDistances}}
\label{sec.proofOfTheorem3}

Recall that we define $d_{AB} = -\frac{3}{4}\log\left(1 - \frac{4}{3}\mathbb{E}\left[\widehat{p}_{AB}\right]\right)$ and Theorem~\ref{thm.nonultrametricDistances}, which we will prove now, tells us that these distances form an additive metric with respect to $S$. %\footnote{Recall that this means that the four point condition holds, i.e., for a quadruple of leaves $A,B,C,D$ that are such that $((A,B),(C,D))$ or $(((A,B),C),D)$ with respect to $S$, the distances satisfy $d_{AB} + d_{CD} \leq d_{AC} + d_{BD} = d_{AD} + d_{BC}$}.  
\begin{theorem3}
The set of dissimilarities $\{d_{AB}\}_{A,B\in L}$ forms an additive metric with respect to $S$. In fact, suppose the leaves $A,B,C,D \in L$ are such that either $((A,B),(C,D))$ or $(((A,B),C),D)$ holds with respect to $S$, then 
\begin{align*}
d_{AC} + d_{BD} &= d_{AD} + d_{BC}> d_{AB} + d_{CD} + \alpha_{\rm add},
\end{align*}
where $\alpha_{\rm add} = \frac{3}{4}\log\left(\frac{8}{3}\mu_L(1-e^{-f}) + 1\right)>0$. 
\end{theorem3}
\begin{IEEEproof}
We will first show that for any 4 leaves $A,B,C,D\in L$ that are such that either $((A,B),(C,D))$ or $(((A,B),C,D))$ holds with respect to $S$,  then $d_{AC} + d_{BD} > d_{AB} + d_{CD} + \alpha_{\rm add}$. Using similar techniques, we will next establish that $d_{AC} + d_{BD} = d_{AB} + d_{CD}$. 

We begin by observing that by definition, 
\begin{align}
d_{AC} + d_{BD} - d_{AB} - d_{CD} &= -\frac{3}{4}\log\left(1 - \frac{4}{3}\mathbb{E}\left[\widehat{p}_{AC}\right]\right) -\frac{3}{4}\log\left(1 - \frac{4}{3}\mathbb{E}\left[\widehat{p}_{BD}\right]\right)\nonumber\\
&\qquad +\frac{3}{4}\log\left(1 - \frac{4}{3}\mathbb{E}\left[\widehat{p}_{AB}\right]\right) + \frac{3}{4}\log\left(1 - \frac{4}{3}\mathbb{E}\left[\widehat{p}_{AB}\right]\right)\\
&= \frac{3}{4}\log\left(\frac{\mathbb{E}\left[e^{-\frac{4}{3}\delta_{AB}}\right]\mathbb{E}\left[e^{-\frac{4}{3}\delta_{CD}}\right]}{\mathbb{E}\left[e^{-\frac{4}{3}\delta_{AC}}\right]\mathbb{E}\left[e^{-\frac{4}{3}\delta_{BD}}\right]}\right),\label{eq.4pointCondition}
\end{align}¥
where the expectations in the last equation are with respect to the multispecies coalescent and the $\delta$'s are the random gene tree distances as defined in Section~\ref{sec.observationModel}. 

We will prove this theorem by lower bounding the quantity $\frac{\mathbb{E}\left[e^{-\frac{4}{3}\delta_{AB}}\right]\mathbb{E}\left[e^{-\frac{4}{3}\delta_{CD}}\right]}{\mathbb{E}\left[e^{-\frac{4}{3}\delta_{AC}}\right]\mathbb{E}\left[e^{-\frac{4}{3}\delta_{BD}}\right]}$ appropriately. Towards this end, we note that for any 4 leaves of the species tree $A,B,C,D$, there are only 2 possible topologies with respect to $S$ upto relabeling: (a) $((A,B),(C,D))$ and (b) $(((A,B),C),D)$. We will consider each case separately and bound the above quantity in what follows. \\\\
\
\underline{Case (a): $((A,B),(C,D))$ } In order to tackle the first case, we will use the notation from Figure~\ref{fig.casea} below, which shows the species tree $S$ restricted to the leaves $A,B,C,D$. Let $o_1, o_2$ and $o_3$ be the common ancestors of $(A,B)$, $(C,D)$ and $(A,C)$ respectively. Let $\mathcal{E}_{AB}$ be the event that the lineages corresponding to $A$ and $B$ coalesce in the segment $(o_1,o_3)$ of the tree in Figure~\ref{fig.casea} and let $\overline{\mathcal{E}_{AB}}$ be the event that this does not occur. Similarly, we define the events $\mathcal{E}_{CD}$ and $\overline{\mathcal{E}_{CD}}$. To reduce notational clutter, for $w,v\in S$, we will write $\mu_{wv}$ to denote $\sum_{e\in\pi_{wv}^S}\mu_e\tau_e$. Now, for leaves $X,Y\in L$, let $Z_{XY}$ denote the random quantity $\frac{1}{2}(\delta_{XY} - \mu_{XY})$, i.e., it is the effective (mutation rate adjusted) coalescent time after the lineages corresponding to $X$ and $Y$ find themselves in a common population.  

By the memoryless property of the exponential distribution, it is easy to check that $Z_{AB} - \mu_{o_1o_3}$ conditioned on $\overline{\mathcal{E}_{AB}}$ has the same distribution as $Z_{CD} - \mu_{o_2o_3}$ conditioned on $\overline{\mathcal{E}_{CD}}$. Let $Z$ denote be a random variable with this common distribution. Also observe that $Z_{AC}$ and $Z_{BD}$ have the same distribution as $Z$. This is depicted diagrammatically in Figure~\ref{fig.casea}. 

Now, using the fact that by definition, $\delta_{AB} = \mu_{AB} + 2Z_{AB}$, we have 
\begin{align}
\mathbb{E}\left[e^{-\frac{4}{3}\delta_{AB}}\right] & = e^{-\frac{4}{3}\mu_{AB}}\mathbb{E}\left[e^{-\frac{8}{3}Z_{AB}}\right]\nonumber\\
&= e^{-\frac{4}{3}\mu_{AB}}\left\{\mathbb{E}\left[e^{-\frac{8}{3}Z_{AB}}\middle | \mathcal{E}_{AB}\right]\mathbb{P}\left(\mathcal{E}_{AB}\right) + \mathbb{E}\left[e^{-\frac{8}{3}Z_{AB}}\middle | \overline{\mathcal{E}_{AB}}\right]\mathbb{P}\left(\overline{\mathcal{E}_{AB}}\right)\right\} \nonumber\\
&\stackrel{(a)}{\geq} e^{-\frac{4}{3}\mu_{AB}}\left\{e^{-\frac{8}{3}\mu_{o_1o_3}}\mathbb{P}\left(\mathcal{E}_{AB}\right) + e^{-\frac{8}{3}\mu_{o_1o_3}}\mathbb{E}\left[e^{-\frac{8}{3}Z}\right]\mathbb{P}\left(\overline{\mathcal{E}_{AB}}\right)\right\}\nonumber\\
& = e^{-\frac{4}{3}\left(\mu_{AB} + 2\mu_{o_1o_3}\right)}\left\{\mathbb{P}\left(\mathcal{E}_{AB}\right) + \mathbb{E}\left[e^{-\frac{8}{3}Z}\right]\mathbb{P}\left(\overline{\mathcal{E}_{AB}}\right)\right\}\label{eq.ABterm},
\end{align}
where $(a)$ follows from the fact that conditioned on $\mathcal{E}_{AB}$, $Z_{AB}\leq \mu_{o_1o_3}$ and that conditioned on $\overline{\mathcal{E}_{AB}}$, $Z_{AB} \stackrel{d}{=} Z + \mu_{o_1o_3}$. Similarly, we get the following lower bound corresponding to the leaves $C,D$. 
% and $\delta_{CD} = \mu_{CD} + 2Z_{CD}$, and the fact that conditioned on $\mathcal{E}_{AB}$ (resp. $\mathcal{E}_{CD}$), $Z_{AB}\leq \mu_{o_1o_3}$ (resp. $Z_{CD}\leq \mu_{o_2o_3}$) to obtain the lower bounds
\begin{align}
%\mathbb{E}\left[e^{-\frac{4}{3}\delta_{AB}}\right]
%&\geq e^{-\frac{4}{3}\mu_{AB}}\Bigg\{ \mathbb{E}\left[e^{-\frac{8}{3}\mu_{o_1o_3}}\middle| \mathcal{E}_{AB}\right]\mathbb{P}\left(\mathcal{E}_{AB}\right)+ e^{-\frac{8}{3}\mu_{o_1o_3}}\mathbb{E}\left[e^{-\frac{8}{3}Z}\right]\mathbb{P}\left(\overline{\mathcal{E}_{AB}}\right)\Bigg\}\label{eq.AB}\\
\mathbb{E}\left[e^{-\frac{4}{3}\delta_{CD}}\right]
&\geq e^{-\frac{4}{3}\left(\mu_{CD} + 2\mu_{o_2o_3}\right)}\left\{\mathbb{P}\left(\mathcal{E}_{CD}\right)+ \mathbb{E}\left[e^{-\frac{8}{3}Z}\right]\mathbb{P}\left(\overline{\mathcal{E}_{CD}}\right)\right\}\label{eq.CDterm}
\end{align}
On the other hand, notice that $\delta_{AC} = \mu_{AC} + 2Z_{AC} \stackrel{d}{=} \mu_{AC} + 2Z$ and $\delta_{BD} = \mu_{BD} + 2Z_{BD} \stackrel{d}{=} \mu_{BD} + 2Z$. Therefore, we have 
\begin{equation}
\mathbb{E}\left[e^{-\frac{4}{3}\delta_{AC}}\right] = e^{-\frac{4}{3}\mu_{AC}}\mathbb{E}\left[e^{-\frac{8}{3}Z}\right], \qquad\mbox{and}\qquad \mathbb{E}\left[e^{-\frac{4}{3}\delta_{BD}}\right] = e^{-\frac{4}{3}\mu_{BD}}\mathbb{E}\left[e^{-\frac{8}{3}Z}\right]\label{eq.ACBDterm},
\end{equation}¥
From equations~\eqref{eq.ABterm} - \eqref{eq.ACBDterm}, we have 
\begin{align}
\frac{\mathbb{E}\left[e^{-\frac{4}{3}\delta_{AB}}\right]}{\mathbb{E}\left[e^{-\frac{4}{3}\delta_{AC}}\right]}&\times\frac{\mathbb{E}\left[e^{-\frac{4}{3}\delta_{CD}}\right]}{\mathbb{E}\left[e^{-\frac{4}{3}\delta_{BD}}\right]}\nonumber\\
& \geq \frac{e^{-\frac{4}{3}\left(\mu_{AB} + 2\mu_{o_1o_3}\right)}\left\{\mathbb{P}\left(\mathcal{E}_{AB}\right) + \mathbb{E}\left[e^{-\frac{8}{3}Z}\right]\mathbb{P}\left(\overline{\mathcal{E}_{AB}}\right)\right\}}{e^{-\frac{4}{3}\mu_{AC}}\mathbb{E}\left[e^{-\frac{8}{3}Z}\right]}\nonumber\\
&\qquad\qquad\times \frac{e^{-\frac{4}{3}\left(\mu_{CD} + 2\mu_{o_2o_3}\right)}\left\{\mathbb{P}\left(\mathcal{E}_{CD}\right) + \mathbb{E}\left[e^{-\frac{8}{3}Z}\right]\mathbb{P}\left(\overline{\mathcal{E}_{AB}}\right)\right\}}{e^{-\frac{4}{3}\mu_{BD}}\mathbb{E}\left[e^{-\frac{8}{3}Z}\right]}\\
&\stackrel{(a)}{=} \frac{\left\{\mathbb{P}\left(\mathcal{E}_{AB}\right) + \mathbb{E}\left[e^{-\frac{8}{3}Z}\right]\mathbb{P}\left(\overline{\mathcal{E}_{AB}}\right)\right\}\left\{\mathbb{P}\left(\mathcal{E}_{CD}\right) + \mathbb{E}\left[e^{-\frac{8}{3}Z}\right]\mathbb{P}\left(\overline{\mathcal{E}_{CD}}\right)\right\}}{\left(\mathbb{E}\left[e^{-\frac{8}{3}Z}\right]\right)^2}\nonumber\\
%&\frac{e^{-\frac{4}{3}\mu_{AB}}\Bigg\{ \mathbb{E}\left[e^{-\frac{8}{3}\mu_{o_1o_3}}\middle| \mathcal{E}_{AB}\right]\mathbb{P}\left(\mathcal{E}_{AB}\right)+ e^{-\frac{8}{3}\mu_{o_1o_3}}\mathbb{E}\left[e^{-\frac{8}{3}Z}\right]\mathbb{P}\left(\overline{\mathcal{E}_{AB}}\right)\Bigg\}}{e^{-\frac{4}{3}\mu_{AC}}\mathbb{E}\left[e^{-\frac{8}{3}Z}\right]}\nonumber\\
%&\qquad\qquad\times \frac{e^{-\frac{4}{3}\mu_{CD}}\Bigg\{ \mathbb{E}\left[e^{-\frac{8}{3}\mu_{o_2o_3}}\middle| \mathcal{E}_{CD}\right]\mathbb{P}\left(\mathcal{E}_{CD}\right)+ e^{-\frac{8}{3}\mu_{o_2o_3}}\mathbb{E}\left[e^{-\frac{8}{3}Z}\right]\mathbb{P}\left(\overline{\mathcal{E}_{C}}\right)\Bigg\}}{e^{-\frac{4}{3}\mu_{BD}}\mathbb{E}\left[e^{-\frac{8}{3}Z}\right]}\nonumber\\
& = \left[\frac{\mathbb{P}\left(\mathcal{E}_{AB}\right)}{\mathbb{E}\left[e^{-\frac{8}{3}Z}\right]} + \mathbb{P}\left(\overline{\mathcal{E}_{AB}}\right)\right]\times\left[\frac{\mathbb{P}\left(\mathcal{E}_{CD}\right)}{\mathbb{E}\left[e^{-\frac{8}{3}Z}\right]} + \mathbb{P}\left(\overline{\mathcal{E}_{CD}}\right)\right]\label{eq.caseARatio}
%\nonumber\\
%&\geq\left[\frac{8}{3}\mu_L\left(1-e^{-f}\right) + 1\right]^2,\label{eq.theorem3a}
 \end{align}
where in $(a)$, we have used the fact that $\mu_{AB} + \mu_{CD} + 2\mu_{o_1o_3} + 2\mu_{o_2o_3} = \mu_{AC} + \mu_{BD}$ and in the last step we divide each term in the numerator by $\mathbb{E}\left[e^{-\frac{8}{3}Z}\right]$. 

Next, observe that $Z$ stochastically dominates the random variable $\mu_L\tilde{Z}$, where $\tilde{Z}\sim $ Exp$(1)$. Therefore, we have 
\begin{equation}
\mathbb{E}\left[e^{-\frac{8}{3}Z}\right]\leq \mathbb{E}\left[e^{-\frac{8}{3}\mu_L\tilde{Z}}\right] = \frac{1}{\frac{8}{3}\mu_L + 1}.
\end{equation}
Substituting this in \eqref{eq.caseARatio} gives us 
\begin{align}
\frac{\mathbb{E}\left[e^{-\frac{4}{3}\delta_{AB}}\right]\mathbb{E}\left[e^{-\frac{4}{3}\delta_{CD}}\right]}{\mathbb{E}\left[e^{-\frac{4}{3}\delta_{AC}}\right]\mathbb{E}\left[e^{-\frac{4}{3}\delta_{BD}}\right]} %&\geq \left[\left(\frac{8}{3}\mu_L + 1\right)\mathbb{P}\left(\mathcal{E}_{AB}\right) + \mathbb{P}\left(\overline{\mathcal{E}_{AB}}\right)\right]\times\left[\left(\frac{8}{3}\mu_L + 1\right)\mathbb{P}\left(\mathcal{E}_{CD}\right)+ \mathbb{P}\left(\overline{\mathcal{E}_{CD}}\right)\right]\nonumber\\
& \geq \left[\frac{8}{3}\mu_L\mathbb{P}\left(\mathcal{E}_{AB}\right) + 1\right]\times\left[\frac{8}{3}\mu_L\mathbb{P}\left(\mathcal{E}_{CD}\right) + 1\right]\label{eq.caseAratioPenultimate}
\end{align}
Finally, we observe that the probability that the event $\mathcal{E}_{AB}$ occurs is given by $ 1 - e^{-\tau_{o_1o_3}}$, where $\tau_{o_1o_3}$ is the length of the path $(o_1,o_3)$ in the species tree; this follows from the memoryless property of the exponential distribution. Since $\tau_{o_1o_3}\geq f$, we have that $\mathbb{P}\left(\mathcal{E}_{AB}\right) \geq 1 - e^{-f}$, and similarly $\mathbb{P}\left(\mathcal{E}_{CD}\right) \geq 1 - e^{-f}$. Substituting this in \eqref{eq.caseAratioPenultimate}, we get the following lower bound
\begin{align}
\frac{\mathbb{E}\left[e^{-\frac{4}{3}\delta_{AB}}\right]\mathbb{E}\left[e^{-\frac{4}{3}\delta_{CD}}\right]}{\mathbb{E}\left[e^{-\frac{4}{3}\delta_{AC}}\right]\mathbb{E}\left[e^{-\frac{4}{3}\delta_{BD}}\right]} &\geq \left[\frac{8}{3}\mu_L\left(1 - e^{-f}\right) + 1\right]^2\label{eq.caseAratioFinal}
\end{align}
%
%
%the last step we have used the fact that the random variable $Z$ stochastically dominates the random variable $\mu_L \tilde{Z}$, where $\tilde{Z}\sim $ Exp$(1)$ and that that 

%$\mathbb{P}\left[\mathcal{E}_{XY}\right]\geq 1 - e^{-f}$ for each pair of leaves $X,Y\in L$.  
\begin{figure*}[!t]
\subfloat[Case (a)]
{\includegraphics[width=0.35\textwidth]{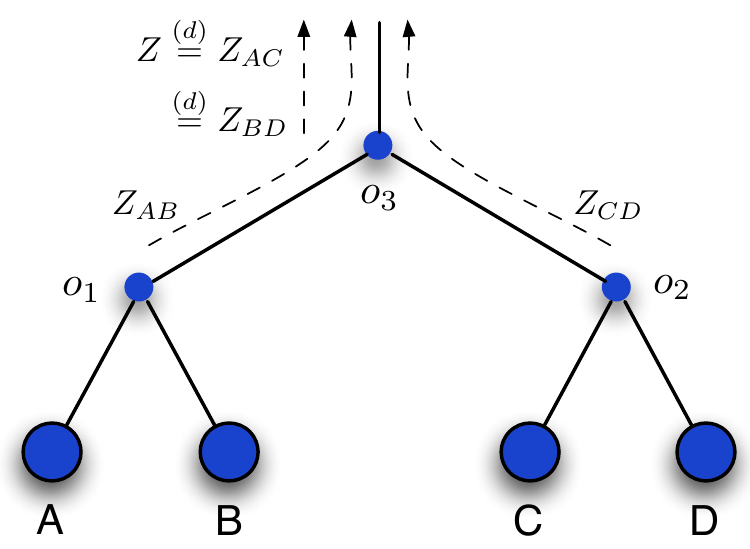}
\label{fig.casea}}
\hfill
\subfloat[Case (b)]
{\includegraphics[width=0.35\textwidth]{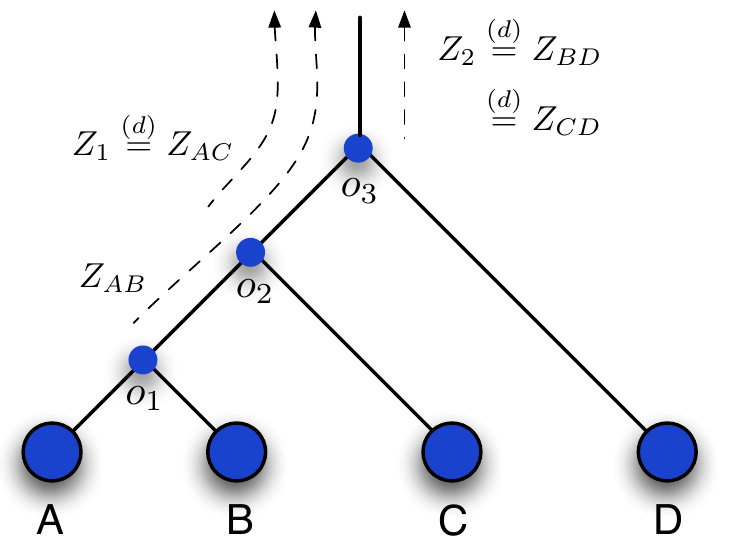}
\label{fig.caseb}}
\caption{Pictures showing the random variables and internal nodes used in Proof of Theorem~\ref{thm.nonultrametricDistances}}
\label{fig.nonultrametricProof}
\end{figure*}
\noindent Next, we consider Case (b). \\\\
\noindent \underline{Case (b) : $(((A,B),C),D)$} Here, we will write $o_1, o_2, o_3$ to denote the most recent common ancestors of $(A,B)$, $(A,C)$ and $(A,D)$ respectively. Again we will use notation from the previous case for random variables of the form $Z_{XY}, X,Y\in L$. In this case, we let $\mathcal{E}_{AB}$ denote the event that the lineages corresponding to $A$ and $B$ coalesce in the branch $(o_1,o_2)$ in Figure~\ref{fig.caseb}. Again, from the memoryless property, it can be seen that  the random variable $Z_{AB} - \mu_{o_1o_2}$ conditioned on $\overline{\mathcal{E}_{AB}}$ and the random variable $Z_{AC}$ have the same distribution; we let $Z_1$ denote a random variable with this common distribution. Similarly  $Z_{CD}$ and $Z_{BD}$ have the same distribution and we let $Z_2$ denote a random variable with this distribution. 

Reasoning as before, we see that since $\delta_{AB} = \mu_{AB} + 2Z_{AB}$, 
\begin{align}
\mathbb{E}\left[e^{-\frac{4}{3}\delta_{AB}}\right] &= e^{-\frac{4}{3}\mu_{AB}}\left\{\mathbb{E}\left[e^{-\frac{8}{3}Z_{AB}}\middle | \mathcal{E}_{AB}\right]\mathbb{P}\left(\mathcal{E}_{AB}\right) + \mathbb{E}\left[e^{-\frac{8}{3}Z_{AB}}\middle | \overline{\mathcal{E}_{AB}}\right]\mathbb{P}\left(\overline{\mathcal{E}_{AB}}\right)\right\} \nonumber\\
&\stackrel{(a)}{\geq} e^{-\frac{4}{3}\mu_{AB}}\left\{e^{-\frac{8}{3}\mu_{o_1o_2}}\mathbb{P}\left(\mathcal{E}_{AB}\right) + e^{-\frac{8}{3}\mu_{o_1o_2}}\mathbb{E}\left[e^{-\frac{8}{3}Z_1}\right]\mathbb{P}\left(\overline{\mathcal{E}_{AB}}\right)\right\}\nonumber\\
& = e^{-\frac{4}{3}\left(\mu_{AB} + 2\mu_{o_1o_2}\right)}\left\{\mathbb{P}\left(\mathcal{E}_{AB}\right) + \mathbb{E}\left[e^{-\frac{8}{3}Z_1}\right]\mathbb{P}\left(\overline{\mathcal{E}_{AB}}\right)\right\}\label{eq.ABtermcaseB},
\end{align}
where, as before, $(a)$ follows from the fact that conditioned on $\mathcal{E}_{AB}$, $Z_{AB}\leq \mu_{o_1o_2}$ and that conditioned on $\overline{\mathcal{E}_{AB}}$, $Z_{AB} \stackrel{d}{=} Z_1 + \mu_{o_1o_2}$. 
On the other hand, we have 
\begin{align}
\mathbb{E}\left[e^{-\frac{4}{3}\delta_{CD}}\right] = e^{-\frac{4}{3}\mu_{CD}}\mathbb{E}\left[e^{-\frac{8}{3}Z_2}\right]\\
\mathbb{E}\left[e^{-\frac{4}{3}\delta_{AC}}\right] = e^{-\frac{4}{3}\mu_{AC}}\mathbb{E}\left[e^{-\frac{8}{3}Z_1}\right]\\
\mathbb{E}\left[e^{-\frac{4}{3}\delta_{BD}}\right] = e^{-\frac{4}{3}\mu_{BD}}\mathbb{E}\left[e^{-\frac{8}{3}Z_2}\right].\label{eq.BDtermcaseB}
\end{align}¥
Therefore, from \eqref{eq.ABtermcaseB}-\eqref{eq.BDtermcaseB}, we have that 
\begin{align}
\frac{\mathbb{E}\left[e^{-\frac{4}{3}\delta_{AB}}\right]\mathbb{E}\left[e^{-\frac{4}{3}\delta_{CD}}\right]}{\mathbb{E}\left[e^{-\frac{4}{3}\delta_{AC}}\right]\mathbb{E}\left[e^{-\frac{4}{3}\delta_{BD}}\right]}&\geq e^{-\frac{4}{3}\left(\mu_{AB} + \mu_{CD} +2\mu_{o_1o_2} - \mu_{AC}-\mu_{BD}\right)}\left(\frac{1}{\mathbb{E}\left[e^{-\frac{8}{3}Z_1}\right]}\mathbb{P}\left(\mathcal{E}_{AB}\right) + \mathbb{P}\left(\overline{\mathcal{E}_{AB}}\right)\right)\nonumber\\
%&= \frac{\mathbb{E}\left[e^{\frac{8}{3}(\mu_{o_1o_2} - Z_{AB})}\middle|\mathcal{E}_{AB}\right]}{\mathbb{E}\left[e^{-\frac{8}{3}Z_1}\right]}\mathbb{P}\left[\mathcal{E}_{AB}\right] + \mathbb{P}\left[\overline{\mathcal{E}_{AB}}\right]\nonumber\\
&= \frac{\mathbb{P}\left[\mathcal{E}_{AB}\right]}{\mathbb{E}\left[e^{-\frac{8}{3}Z_1}\right]} + \mathbb{P}\left[\overline{\mathcal{E}_{AB}}\right]\label{eq.caseBratioPenultimate}
\end{align}
where the second step follows from the fact that $\mu_{AB} + \mu_{CD} +2\mu_{o_1o_2} = \mu_{AC}+\mu_{BD}$. Finally, as in case (a), we use the bounds $\mathbb{E}\left[e^{-\frac{8}{3}Z_1}\right]\leq \frac{1}{\frac{8}{3}\mu_L +1}$ and that $\mathbb{P}\left[\mathcal{E}_{AB}\right] \geq 1 - e^{-f}$ to get the following lower bound.  
%for all pairs of leaves $X,Y\in L$. Taking logarithms on either side of \eqref{eq.theorem3a} and \eqref{eq.theorem3b} gives us the result that $d_{AC} + d_{BD}\geq d_{AB} + d_{CD} + \log\left(\frac{8}{3}\mu_L\left(1-e^{-f}\right) + 1\right)$.  Using a similar procedure, one can show that $d_{AC} + d_{BD} = d_{AD} + d_{BC}$. This proves that $\{d_{AB}\}_{A,B\in L}$ is an additive metric with respect to $S$. 
\begin{equation}
\frac{\mathbb{E}\left[e^{-\frac{4}{3}\delta_{AB}}\right]\mathbb{E}\left[e^{-\frac{4}{3}\delta_{CD}}\right]}{\mathbb{E}\left[e^{-\frac{4}{3}\delta_{AC}}\right]\mathbb{E}\left[e^{-\frac{4}{3}\delta_{BD}}\right]} \geq \frac{8}{3}\mu_L(1 - e^{-f}) + 1\label{eq.caseBratioFinal}
\end{equation}¥

Since $\left(\frac{8}{3}\mu_L(1 - e^{-f}) + 1\right)\geq 1$, from \eqref{eq.caseAratioFinal} and \eqref{eq.caseBratioFinal}, we have that for any 4 leaves $A,B,C,D$ such that the species tree $S$ restricted to these four leaves satisfies either $((A,B),(C,D))$ or $(((A,B),C),D)$, then
\begin{equation}
\frac{\mathbb{E}\left[e^{-\frac{4}{3}\delta_{AB}}\right]\mathbb{E}\left[e^{-\frac{4}{3}\delta_{CD}}\right]}{\mathbb{E}\left[e^{-\frac{4}{3}\delta_{AC}}\right]\mathbb{E}\left[e^{-\frac{4}{3}\delta_{BD}}\right]} \geq \frac{8}{3}\mu_L(1 - e^{-f}) + 1
\end{equation}
Substituting this lower bound in \eqref{eq.4pointCondition}, we get the  result that for any 4 leaves $A,B,C,D\in L$ that are such that $((A,B),(C,D))$ or $(((A,B),C,D))$ holds with respect to $S$, we have that $d_{AC} + d_{BD} > d_{AB} + d_{CD} + \alpha_{\rm add}$, where $\alpha_{\rm add} = \frac{3}{4}\log\left(\frac{8}{3}\mu_L(1-e^{-f}) + 1\right)$.

To conclude the proof, we will next establish the ``equality part'' of the theorem. As in \eqref{eq.4pointCondition}, notice that the following holds.  
\begin{align}
d_{AC} + d_{BD} - d_{AD} - d_{BC} &= \frac{3}{4}\log\left(\frac{\mathbb{E}\left[e^{-\frac{4}{3} \delta_{AD}}\right]\mathbb{E}\left[e^{-\frac{4}{3} \delta_{BC}}\right]}{\mathbb{E}\left[e^{-\frac{4}{3} \delta_{AC}}\right]\mathbb{E}\left[e^{-\frac{4}{3} \delta_{BD}}\right]}\right)\label{eq.equality}.
\end{align}
Again, we will divide this proof into two cases as above. 

\noindent\underline{Case (a): $((A,B),(C,D))$ }  Observe that the following hold with $\mu_{XY}$ and $Z$ as defined before (cf. Fig~\ref{fig.casea})
\begin{align*}
\delta_{AD} &= \mu_{AD} + 2Z,\;\;
\delta_{BC} = \mu_{BC} + 2Z,\\
\delta_{AC} &= \mu_{AC} + 2Z,\;\;
\delta_{BD} = \mu_{BD} + 2Z
\end{align*}
Substituting these in \eqref{eq.equality} and observing that $\mu_{AD} + \mu_{BC} = \mu_{AC} + \mu_{BD}$ tells us that $d_{AC} + d_{BD} = d_{AD} + d_{BC}$ in case (a). 

\noindent\underline{Case (b): $(((A,B),C),D)$} In this case, observe that the following hold again with $\mu_{XY}$ and $Z_1$ and $Z_2$ as defined earlier (cf. Fig 2(b)): 
\begin{align*}
\delta_{AD} &= \mu_{AD} + 2Z_2,\;\;
\delta_{BC} = \mu_{BC} + 2Z_1\\
\delta_{AC} &= \mu_{AC} + 2Z_1,\;\;
\delta_{BD} = \mu_{BD} + 2Z_2
\end{align*}
Again, substituting these in \eqref{eq.equality} and observing that $\mu_{AD} + \mu_{BC} = \mu_{AC} + \mu_{BD}$ tells us that $d_{AC} + d_{BD} = d_{AD} + d_{BC}$ in case (b) as well. This concludes the proof.
%\pagebreak

\end{IEEEproof}

%%%%%%%%%%%%%%%%%%%%%%%%%%%%%%%%%%%%%
\section{Proof of Theorem~\ref{thm.nonultrametricSampleComplexity}}
\label{sec.proofOfTheorem4}
%%%%%%%%%%%%%%%%%%%%%%%%%%%%%%%%%%%%%
We will now prove the last main result in our paper that shows that Theorem~\ref{thm.nonultrametricDistances} can be used to design a tree reconstruction algorithm when one only has access to molecular data and also provides sample complexity results for this algorithm. Recall that we propose   the following measure of dissimilarity from the samples
\begin{align}
\widehat{d}_{AB} &\triangleq-\frac{3}{4}\log\left(1 - \frac{4}{3}\,\widehat{p}_{AB}\right).
\end{align}
where $\widehat{p}_{AB}$ is as defined in \eqref{eq.SampleUltrametricDistances}. 

In light of Theorem~\ref{thm.nonultrametricDistances}, we proposed the following tree reconstruction procedure, which we call METAL: use any distance algorithm (like Neighbor Joining \cite{saitou1987neighbor}) which returns an additive tree using $\{\widehat{d}_{AB}\}_{A,B\in L}$ as the dissimilarity measure. We then have the following result. 
\begin{theorem4}
For any $\epsilon >0$, the METAL algorithm succeeds in reconstructing (the unrooted version of) $S$ with probability at least $1 - \epsilon$ as long as $m$ and $k$ satisfy
\begin{align}
k\geq 1\,\mbox{and}\,\,m &\geq \frac{e^{\frac{8\mu_U \Delta}{3}}({8}\mu_U+3)^2(24+8\alpha_{\rm add})^2 }{162\alpha_{\rm add}^2}\log\left(\frac{16{n\choose 4}}{\epsilon}\right)\label{eq.thm4SampleComplexityAppendix}
\end{align}
where $\alpha_{\rm add} = \frac{3}{4}\log\left(\frac{8}{3}\mu_L(1-e^{-f}) + 1\right)$. 

In the limit as $f\to 0$, the right side above approaches $$C_2(\mu_U,\mu_L,\Delta,n,\epsilon)~\times~f^{-2}, \mbox{ where } C_2(\mu_U,\mu_L,\Delta,n,\epsilon) = \frac{8e^{\frac{8\mu_U \Delta}{3}}(8\mu_U+3)^2 }{9\mu_L^2}\log\left(\frac{16{n\choose 3}}{\epsilon}\right).$$ 
\end{theorem4}
\begin{IEEEproof}
Notice that the above algorithm makes an error only if there exists a set of four leaves $A,B,C,D$ such that $\tau_{AB} + \tau_{CD} \leq \tau_{AC} + \tau_{BD} = \tau_{AD} + \tau_{BC}$, but the 4-point condition is not satisfied by $\widehat{d}$, that is: 
$$\widehat{d}_{AB} + \widehat{d}_{CD} - \widehat{d}_{AC} - \widehat{d}_{BD} > 0\;\; \mbox{ or }\;\;\widehat{d}_{AB} + \widehat{d}_{CD} - \widehat{d}_{AD} - \widehat{d}_{BC} > 0$$
 Therefore, using the union bound, the probability of error can be upper bounded as follows:
\begin{align}
\mathbb{P}\left(\mbox{Error}\right) &\leq \sum_{\substack{A,B,C,D\in L:\\\tau_{AB} + \tau_{CD}\leq \tau_{AC} + \tau_{BD} = \tau_{AD} + \tau_{BC}}} \mathbb{P}\left[\hat{d}_{AB} + \hat{d}_{CD} - \hat{d}_{AC} - \hat{d}_{BD} > 0\right] \nonumber\\&\qquad\qquad\qquad\qquad\qquad\qquad\qquad+ \mathbb{P}\left[\hat{d}_{AB} + \hat{d}_{CD} - \hat{d}_{AD} - \hat{d}_{BC} > 0\right]\label{eq.thm4PError}
\end{align} 

We will bound the first term inside the summation of \eqref{eq.thm4PError} and the second one will follow similarly. Setting $\alpha_{\rm add} \triangleq \frac{3}{4}\log\left(\frac{8}{3}\mu_L(1-e^{- f}) + 1\right)$, observe that for a quadruple of leaves $A,B,C,D$ such that $\tau_{AB} + \tau_{CD} \leq \tau_{AC} + \tau_{BD} = \tau_{AD} + \tau_{BC}$, we have 
\begin{align*}
\mathbb{P}\left[\widehat{d}_{AB} + \widehat{d}_{CD} - \widehat{d}_{AC} - \widehat{d}_{BD} > 0\right] 
& = \mathbb{P}\Big[\widehat{d}_{AB} - d_{AB} + \widehat{d}_{CD} - d_{CD}- \widehat{d}_{AC} + d_{AC} - \widehat{d}_{BD} + d_{BD} \Big.\\
&\qquad\Big.> d_{AC} + d_{BD} - d_{AB}- d_{CD}\Big] \\
&\leq \mathbb{P}\Big[\widehat{d}_{AB} - d_{AB} + \widehat{d}_{CD} - d_{CD}- \widehat{d}_{AC} + d_{AC} - \widehat{d}_{BD} + d_{BD} > \alpha_{\rm add}\Big],
\end{align*}
where the second inequality follows from Theorem~\ref{thm.nonultrametricDistances} which says that $d_{AC} + d_{BD} - d_{AB}- d_{CD} > \alpha_{\rm add}$. We will again use the union bound to get
\begin{align}
\mathbb{P}\left[\widehat{d}_{AB} + \widehat{d}_{CD} - \widehat{d}_{AC} - \widehat{d}_{BD} > 0\right] &\leq \mathbb{P}\Big[\widehat{d}_{AB} - d_{AB} + \widehat{d}_{CD} - d_{CD}- \widehat{d}_{AC} + d_{AC} - \widehat{d}_{BD} + d_{BD} > \alpha_{\rm add}\Big]\nonumber\\
&\leq \mathbb{P}\left[\widehat{d}_{AB} - d_{AB} > \frac{\alpha_{\rm add}}{4}\right] + \mathbb{P}\left[\widehat{d}_{CD} - d_{CD} > \frac{\alpha_{\rm add}}{4}\right]\nonumber\\
&\pushright{+ \mathbb{P}\left[d_{AC}-\widehat{d}_{AC}  > \frac{\alpha_{\rm add}}{4}\right] + \mathbb{P}\left[d_{BD}-\widehat{d}_{BD} > \frac{\alpha_{\rm add}}{4}\right].\;\;\qquad{}}\label{eq.thm4PError2}
\end{align}
To proceed, we will focus our attention on the first term in \eqref{eq.thm4PError2}. The remaining terms will follow similarly. For notational clarity, let us define the function $\ell(x)\triangleq -\frac{3}{4}\log\left(1 - \frac{4}{3}x\right)$ and let $p_{AB}^{(i)}$  denote the random quantity $\frac{3}{4}\left(1 - e^{-\frac{4}{3}\delta^{(i)}_{AB}} \right) = \ell^{-1}\left(\delta_{AB}^{(i)}\right)$, where, as usual, $\delta^{(i)}_{AB}$ is the distances between $A$ and $B$ on the random gene tree $\gcal^{(i)}$  drawn according to the MSC. Now, observe that, by definition, $\widehat{d}_{AB}$ and $d_{AB}$ are equal to $\ell(\widehat{p}_{AB})$ and $\ell(\mathbb{E}\left[\widehat{p}_{AB}\right])$ respectively. 

Our strategy will be to first show that with high probability $\widehat{p}_{AB}$ is close to $\frac{1}{m}\sum_{i=1}^m p_{AB}^{(i)}$ which is in turn close to $\mathbb{E}\left[\widehat{p}_{AB}\right]$. We will then use the fact that $\ell(x)$ is a well-behaved function to obtain an upper bound on the the first term of \eqref{eq.thm4PError2}. 

Conditioned on a particular realization of the MSC process $\left\{\delta^{(i)}_{AB}\right\}  _{i\in[m]}$, let $\mathcal{E}_1(\xi)$ and $\mathcal{E}_2(\xi)$ denote the events that  $\left|\frac{1}{m}\sum_{i\in [m]}{p}^{(i)}_{AB} - \mathbb{E}\widehat{p}_{AB}\right| >\xi$ and $\left|\frac{1}{m}\sum_{i\in [m]}{p}^{(i)}_{AB} - \widehat{p}_{AB}\right| >\xi$, respectively. 
Now, notice we can bound the first term in \eqref{eq.thm4PError2} as follows. 
\begin{align}
\mathbb{P}\left[\widehat{d}_{AB} - d_{AB} > \frac{\alpha_{\rm add}}{4}\right]&= \mathbb{P}\left[\ell(\widehat{p}_{AB}) - \ell(\mathbb{E}\widehat{p}_{AB})> \frac{\alpha_{\rm add}}{4}\right]\nonumber\\
&\stackrel{(a)}{=}\mathbb{E}\Bigg[\mathbb{P}\left[\ell(\widehat{p}_{AB}) - \ell(\mathbb{E}\widehat{p}_{AB})> \frac{\alpha_{\rm add}}{4}\middle | \left\{\delta^{(i)}_{AB}\right\}_{i\in [m]}\right]\Bigg]\nonumber\\
%&\stackrel{(b)}{\leq} \mathbb{E}\left[\mathbb{P}\left(\mathcal{E}_1(\xi)\middle | \left\{\delta^{(i)}_{AB}\right\}_{i\in [m]}\right)\right] + \mathbb{E}\left[\mathbb{P}\left(\mathcal{E}_2(\xi)\middle | \left\{\delta^{(i)}_{AB}\right\}_{i\in [m]}\right)\right] \nonumber \\
%&\pushright{\qquad+\mathbb{E}\left[\mathbb{P}\left[\ell(\widehat{p}_{AB}) - \ell(\mathbb{E}\widehat{p}_{AB})> \frac{\alpha_{\rm add}}{4}\middle | \left\{\delta^{(i)}_{AB}\right\}_{i\in [m]},\mathcal{E}_1(\xi)^c,\mathcal{E}_2(\xi)^c\right]\right],}\label{eq.conditioning}
&\stackrel{(b)}{\leq} \mathbb{E}\left[\mathbb{P}\left[\ell(\widehat{p}_{AB}) - \ell(\mathbb{E}\widehat{p}_{AB})> \frac{\alpha_{\rm add}}{4}\middle | \left\{\delta^{(i)}_{AB}\right\}_{i\in [m]},\mathcal{E}_1(\xi)^c,\mathcal{E}_2(\xi)^c\right]\right]\nonumber \\
&\pushright{\qquad+\mathbb{E}\left[\mathbb{P}\left(\mathcal{E}_1(\xi)\middle | \left\{\delta^{(i)}_{AB}\right\}_{i\in [m]}\right)\right] + \mathbb{E}\left[\mathbb{P}\left(\mathcal{E}_2(\xi)\middle | \left\{\delta^{(i)}_{AB}\right\}_{i\in [m]}\right)\right],}\label{eq.conditioning}
\end{align}
where in $(a)$ we condition on $\left\{\delta^{(i)}_{AB}\right\}$, a particular realization of the MSC. In $(b)$ we use the following fact: for any three events $\mathcal{E}_a, \mathcal{E}_b,\mathcal{E}_c$, the following inequality holds
\begin{align*}
\mathbb{P}(\mathcal{E}_a) &= \mathbb{P}(\mathcal{E}_a| \mathcal{E}_b\cup \mathcal{E}_c)\mathbb{P}\left(\mathcal{E}_b\cup \mathcal{E}_c\right) + \mathbb{P}(\mathcal{E}_a| \mathcal{E}_b^c\cap \mathcal{E}_c^c)\mathbb{P}\left(\mathcal{E}_b^c\cap \mathcal{E}_c^c\right)\\
&\leq \mathbb{P}(\mathcal{E}_b\cup \mathcal{E}_c) + \mathbb{P}(\mathcal{E}_a| \mathcal{E}_b^c\cap \mathcal{E}_c^c)\\
&\leq \mathbb{P}\left(\mathcal{E}_b\right) + \mathbb{P}\left(\mathcal{E}_c\right) + \mathbb{P}(\mathcal{E}_a| \mathcal{E}_b^c\cap \mathcal{E}_c^c),
\end{align*}
where we identify $\mathcal{E}_a, \mathcal{E}_b$, and $\mathcal{E}_c$ with the events $\widehat{d}_{AB} - d_{AB} > \frac{\alpha_{\rm add}}{4}, \mathcal{E}_1(\xi)$, and $\mathcal{E}_2(\xi)$ respectively. Our goal now is to pick a value of $\xi$ so that the first term in \eqref{eq.conditioning} is 0. Towards this end, we will use the following result that we prove in Section~\ref{sec.proofOfClaim1}. 
\begin{claim}
\label{claim.claim1}
For any $\xi >0$, conditioned on a particular realization $\left\{\delta^{(i)}_{AB}\right\}_{i\in [m]}$ of the MSC process, and the events $\mathcal{E}_1(\xi)^c$ and $\mathcal{E}_2(\xi)^c$, the following inequality holds 
\begin{equation}
\left|\ell(\widehat{p}_{AB}) - \ell(\mathbb{E}\widehat{p}_{AB})\right| \leq \frac{2\xi}{\frac{e^{-4\mu_U \Delta/3}}{\frac{8}{3}\mu_U +1} - \frac{8\xi}{3}}\label{eq.claim1}.
\end{equation}
\end{claim}
%\begin{IEEEproof}
%\textcolor{red}{the proof goes here}
%\end{IEEEproof}

Now, Claim~\ref{claim.claim1} tells us that if we make the following choice for $\xi$
\begin{equation}
\xi = \xi_0 \triangleq \frac{9 \alpha_{\rm add}e^{-\frac{4}{3}\mu_U\Delta}}{(24 + \alpha_{\rm add})(8\mu_U + 3)}\label{eq.choiceOfKappa},
\end{equation}
then conditioned on the events $\mathcal{E}_1(\xi_0)^c$ and $\mathcal{E}_2(\xi_0)^c$, we have that 
$$\ell(\widehat{p}_{AB}) - \ell(\mathbb{E}\widehat{p}_{AB}) \leq \frac{\alpha_{\rm add}}{4}$$

%we have that the right side of \eqref{eq.claim1} is equal to $\frac{\alpha_{\rm add}}{4}$. 
Therefore, we have 
\begin{equation}
\mathbb{P}\left[\ell(\widehat{p}_{AB}) - \ell(\mathbb{E}\widehat{p}_{AB})> \frac{\alpha_{\rm add}}{4}\middle | \left\{\delta^{(i)}_{AB}\right\}_{i\in [m]},\mathcal{E}_1(\xi_0)^c,\mathcal{E}_2(\xi_0)^c\right] = 0.
\end{equation}
Using this in \eqref{eq.conditioning}, we have 
\begin{align}
\mathbb{P}\left[\widehat{d}_{AB} - d_{AB} > \frac{\alpha_{\rm add}}{4}\right] & \leq \mathbb{E}\left[\mathbb{P}\left(\mathcal{E}_1(\xi_0)\middle | \left\{\delta^{(i)}_{AB}\right\}_{i\in [m]}\right)\right] + \mathbb{E}\left[\mathbb{P}\left(\mathcal{E}_2(\xi_0)\middle | \left\{\delta^{(i)}_{AB}\right\}_{i\in [m]}\right)\right]\nonumber\\
&\leq e^{-2m\xi_0^2} + e^{-2mk\xi_0^2}\label{eq.hoeffdings},
\end{align}
where the second inequality comes from applying Hoeffding's inequality to each term, as in \eqref{eq.thm2Hoeffding1} and \eqref{eq.thm2Hoeffding2}. Since this upper bound is independent of the choice of the pair of leaves, we can use \eqref{eq.hoeffdings} and \eqref{eq.thm4PError2} in \eqref{eq.thm4PError} to get 

\begin{align}
\mathbb{P}\left[\mbox{Error}\right] \leq \sum_{\substack{A,B,C,D\in L:\\\tau_{AB} + \tau_{CD}\leq \tau_{AC} + \tau_{BD} = \tau_{AD} + \tau_{BC}}} 8\left(e^{-2m\xi_0^2} + e^{-2mk\xi_0^2} \right)\nonumber\\
\leq 8{n\choose 4} \left(e^{-2m\xi_0^2} + e^{-2mk\xi_0^2} \right).
\end{align}
Now, if we pick $m$ and $k$ as in \eqref{eq.thm4SampleComplexityAppendix} (also \eqref{eq.thm4SampleComplexity}), we see that the right side above is less than $\epsilon$, which concludes the proof. The limit as $f\to 0$ can also be readily computed by observing that $\alpha_{\rm add}\to 2\mu_L f$ as $f\to 0$.
\end{IEEEproof}
\subsection{Proof of Claim~\ref{claim.claim1}}
\label{sec.proofOfClaim1}

We will begin by using the fact that $\ell(x)$ satisfies the following Lipschitz property: for any $0\leq x\leq y\leq B$, we have
\begin{align}
\ell(y) - \ell(x) &= -\frac{3}{4}\log\left(1 - \frac{4}{3}y\right) + \frac{3}{4}\log\left(1 - \frac{4}{3}x\right)\nonumber\\
&=\int_x^y\frac{1}{1 - \frac{4}{3}t}\,dt\nonumber\\
&\leq \frac{(y-x)}{1 - \frac{4}{3}B}.\label{eq.lipschitz}
\end{align}

From this, we have that 
\begin{align}
\left|\ell\left(\frac{1}{m}\sum_{i\in [m]}{p}^{(i)}_{AB}\right) - \ell\left(\mathbb{E}[p_{AB}]\right)\right|\leq \frac{\xi}{1 - \frac{4}{3}\left(\mathbb{E}\left[\widehat{p}_{AB}\right] + \xi\right)},\;\;\mbox{ conditioned on $\mathcal{E}_1(\xi)$},\label{eq.applyingLipschitz}
\end{align}
where we have chosen the  $B$ (of \eqref{eq.lipschitz}) to be $\mathbb{E}\left[\widehat{p}_{AB}\right] + \xi$, since conditioned on $\mathcal{E}_1(\xi)$, we have that 
\begin{equation}
\frac{1}{m}\sum_{i=1}^mp^{(i)}_{AB}\leq \mathbb{E}\left[\widehat{p}_{AB}\right] + \xi.\label{eq.conditionedonE1}
\end{equation}
Similarly, conditioned on $\mathcal{E}_2(\xi)^c$ and $\mathcal{E}_1(\xi)^c$, we have 
\begin{align}
\left|\ell\left(\frac{1}{m}\sum_{i\in [m]}{p}^{(i)}_{AB}\right) - \ell\left(\widehat{p}_{AB}\right)\right|& \leq \frac{\xi}{1 - \frac{4}{3}\left(\frac{1}{m}\sum_{i\in[m]}p^{(i)}_{AB} + \xi\right)}\nonumber \\
&\leq \frac{\xi}{1 - \frac{4}{3}\left(\mathbb{E}\left[\widehat{p}_{AB}\right] + 2\xi\right)}\label{eq.applyingLipschitz2},
\end{align}
where in the first inequality we have chosen $B$ (of \eqref{eq.lipschitz}) to be $\frac{1}{m}\sum_{i\in[m]}p^{(i)}_{AB} + \xi$, since conditioned on $\mathcal{E}_2(\xi)^c$, we have that $\widehat{p}_{AB}\leq\frac{1}{m}\sum_{i\in[m]}p^{(i)}_{AB} + \xi$,  and the second inequality follows from \eqref{eq.conditionedonE1}. Therefore from \eqref{eq.applyingLipschitz} and \eqref{eq.applyingLipschitz2}, we have that the following inequality holds conditioned on $\mathcal{E}_1(\xi)^c$ and $\mathcal{E}_2(\xi)^c$:
\begin{align}
\left|\ell(\widehat{p}_{AB}) - \ell(\mathbb{E}\left[\widehat{p}_{AB}\right])\right| &\leq \left|\ell\left(\frac{1}{m}\sum_{i\in [m]}{p}^{(i)}_{AB}\right) - \ell\left(\mathbb{E}[p_{AB}]\right)\right|+\left|\ell\left(\frac{1}{m}\sum_{i\in [m]}{p}^{(i)}_{AB}\right) - \ell\left(\widehat{p}_{AB}\right)\right|\nonumber\\
&\leq \frac{2\xi}{1 - \frac{4}{3}\left(\mathbb{E}\left[\widehat{p}_{AB}\right] + 2\xi\right)}\label{eq.needToBoundpAB}
\end{align}

Finally, to conclude the proof of the claim, we bound $\mathbb{E}[p_{AB}]$ using the properties of the multispecies coalescent. Notice that, by definition, the random distance $\delta_{AB}$ is equal to $\mu_{AB} + 2 Z_{AB}$, where $\mu_{AB}$ and $Z_{AB}$ are as defined in Section~\ref{sec.proofOfTheorem3}. Therefore, 
\begin{align}
\mathbb{E}\left[\widehat{p}_{AB}\right] & = \mathbb{E}\left[\frac{3}{4}(1 - e^{-\frac{4}{3}\delta_{AB}})\right]\nonumber\\
& = \frac{3}{4}\left(1 - e^{-\frac{4}{3}\mu_{AB}}\mathbb{E}\left[e^{-\frac{8}{3}Z_{AB}}\right]\right)\label{eq.pABbound}
%&\leq \frac{3}{4}\left(1 - \frac{e^{-\frac{4}{3}\mu_U\Delta}}{\frac{8}{3}\mu_U + 1}\right)\label{eq.expectationUpperBound}.
\end{align} 
Next, we observe that the random variable $Z_{AB}$ is stochastically dominated by the random variable $\mu_UZ$, where $Z\sim$ Exp$(1)$. This implies that 
\begin{align*}
\mathbb{E}\left[e^{-\frac{8}{3}Z_{AB}}\right] &\geq \mathbb{E}\left[e^{-\frac{8}{3}\mu_UZ}\right]\\
& = \frac{1}{\frac{8}{3}\mu_U + 1}.
\end{align*}
Using this and the fact that $\mu_{AB}\leq \mu_U\Delta$ in \eqref{eq.pABbound}, we have 
$$\mathbb{E}\left[\widehat{p}_{AB}\right]\leq \frac{3}{4}\left(1 - \frac{e^{-\frac{4}{3}\mu_U\Delta}}{\frac{8}{3}\mu_U + 1}\right).$$
Substituting this in \eqref{eq.needToBoundpAB} concludes the proof. 
\end{document}